\documentclass[3p,times]{elsarticle}

\usepackage{amsmath,amssymb,amsfonts}
\usepackage{amsthm}
\usepackage{bm}
\usepackage{url}
\usepackage{graphicx}
\usepackage{booktabs}
\usepackage{makecell}
\usepackage{multirow}
\usepackage{xcolor}
\usepackage[inline]{enumitem}
\usepackage{placeins}
\usepackage{capt-of}
\usepackage{wrapfig}
\usepackage[hidelinks]{hyperref}

\newtheorem{theorem}{Theorem}
\newtheorem{proposition}{Proposition}
\newtheorem{lemma}{Lemma}
\newtheorem{corollary}{Corollary}
\theoremstyle{definition}
\newtheorem{definition}{Definition}
\newtheorem{assumption}{Assumption}

\newcommand{\Cg}{\mathcal{C}^{+}}
\newcommand{\Cb}{\mathcal{C}^{-}}
\newcommand{\Ag}{\mathcal{A}^{+}}
\newcommand{\Ab}{\mathcal{A}^{-}}

\makeatletter
\def\ps@pprintTitle{%
  \let\@oddhead\@empty\let\@evenhead\@empty
  \def\@oddfoot{\footnotesize\itshape Preprint version\hfill}%
  \let\@evenfoot\@oddfoot}
\makeatother
\journal{Journal of Network and Computer Applications}

\begin{document}

\begin{frontmatter}

\title{DeFSM: Robust and Efficient Decentralization of Federated Submodular Maximization}

\author{Duc A. Tran}
\author{Dung Truong\corref{cor1}}
\ead{dung.truong001@umb.edu}
\author{Duy Le}
\cortext[cor1]{Corresponding author.}

\affiliation{organization={Department of Computer Science, University of Massachusetts Boston},
            city={Boston},
            state={MA},
            country={USA}}

\begin{abstract}
Submodular maximization seeks a small subset best representing a large dataset. In its federated setting, the server is a single point of failure and some clients may be Byzantine. Removing the server raises a trilemma of decentralization, efficiency, and security.
We propose DeFSM, a Byzantine-robust and communication-efficient architecture that replaces the server with a lightweight smart contract and a pool of independent, consensus-free agents. Its aggregation rule, SafeGM, builds on two notions from combinatorial geometry: Tukey depth and safe points (the points lying in the convex hull of the good points wherever the bad points are). SafeGM makes the robust decision on 2D Gaussian random projections, which the contract can compute efficiently, and lifts it back as a convex combination of actual updates. Since lifting loses information, it uses the projected geometric median only when its depth certifies it, and otherwise a Tukey median.
The share of bad updates is then bounded only by a count: how many there are, not what they contain. We prove, and confirm empirically, that this bound controls the solution quality. On CIFAR10 and PathMNIST, with up to 30\% Byzantine clients and agents, DeFSM never collapses and retains on average about 91\% of the quality of the ideal reference, the geometric median at a trusted central server. The geometric median decentralized in its best setting would retain about 96\%, but the contract cannot compute it, and so each agent would have to download six times more data, or each client twice as much.
\end{abstract}


\end{frontmatter}

\section{Introduction}\label{sec:intro}
Edge devices generate vast datasets that can fuel machine learning for predictive analytics, anomaly detection, and resource optimization. Transmitting all of these data, however, strains bandwidth and the central server, and labeling them requires substantial labor. Since data are often redundant, with additional samples adding less benefit, we should collect only the most informative ones. This is one of many applications of Federated Submodular Maximization (FSM), the federated setting of submodular maximization, a classic problem with diminishing marginal returns.
FSM is analogous to Federated Learning (FL), except that the unknown is a subset and the objective is a submodular set function. Whereas FL has gained great traction, FSM research is rare \cite{DBLP:conf/icml/Rafiey24,10286024,TranBigData2025}.

An FSM system consists of local machines, the \emph{clients}, and a central \emph{aggregation server}. In the basic protocol for learning top-$r$ elements from a federated dataset \cite{DBLP:conf/icml/Rafiey24}, the server broadcasts a fractional global solution $\mathbf{x}$, initially zero. Each client returns, as a binary vector, the $r$ elements it would add under its local objective around $\mathbf{x}$. The server aggregates these vectors, typically by averaging, and moves $\mathbf{x}$ a small step in the aggregated direction. After a fixed number of rounds, $\mathbf{x}$ is rounded to the final subset.

The federated setting requires a central server. In practice, such a server may not be feasible due to limited resources, lack of trust, or the risk of a single point of failure. Also, the solution quality relies on the clients following the protocol, yet they can submit fake local updates.

\textbf{Problem. } We address a new problem: decentralized FSM without a server and in the presence of Byzantine clients. As for decentralization, a smart contract on a blockchain can replace the server logic verifiably, but it can afford only light computation, communication, and storage. The aggregation must therefore be outsourced to off-chain workers, which we call \emph{agents}. However, some agents may themselves be Byzantine. No FSM research exists to date in this direction.
As for client misbehavior, robust aggregators exist for federated learning \cite{pillutla2022robust,Yin2018ByzantineRobustDL,10.5555/3294771.3294783}, but robust FSM is underexplored \cite{TranBigData2025}. How well these aggregators work when decentralized in server-less FSM is unknown.

\textbf{Challenges. }Robust aggregators operate on high-dimensional vectors, which the contract cannot afford, so the agents must run them entirely off-chain. Each agent then downloads all other agents' aggregates in every round, hence incurring substantial traffic. Moreover, the agents need to agree on a final aggregate, but vector consensus in high dimensions is intractable in general \cite{10.1145/2484239.2484256}. Decentralization, efficiency, and security thus form a trilemma.

\textbf{Contributions. }
We propose $\mathsf{DeFSM}$, a decentralized architecture for FSM, with its aggregation rule $\mathsf{SafeGM}$. Together they resolve the trilemma, and no party needs to know how many clients or agents are Byzantine.
The key idea is \emph{safety}: an aggregate is safe if it lies in the convex hull of the good updates. Tukey depth can certify this: a point is safe if its depth exceeds the number of bad points. 
$\mathsf{SafeGM}$ projects the updates onto a random 2D plane and computes their geometric median; if its depth falls below our design threshold, a Tukey median is used instead. This 2D point is then lifted back to the original space $\mathbb{R}^d$ as a convex combination of actual updates, in which bad updates receive a share bounded by their count.
Safety controls the solution quality because of two properties specific to FSM, which federated learning lacks: monotone objectives and sparse binary updates. Our contributions:
\begin{itemize}
\item
\underline{Novel architecture}: $\mathsf{DeFSM}$ is, to our knowledge, the first FSM architecture without a trusted server. Aggregation is outsourced to a pool of independent agents, a role anyone, even a client, can take, and coordinated by a smart contract so light that any blockchain can host it: it stores only hashes, seeds, and a few numbers per agent. Clients share their updates through IPFS, and each round the contract itself, not a vote among agents, makes the robust decision.
\item
\underline{Provable robustness}: Today's robust aggregators judge updates by distances, which an attacker can shape. Our aggregation rule, $\mathsf{SafeGM}$, offers a guarantee that no manipulation of data can break: the share of bad updates in its aggregate is bounded only by how many there are, not by what they contain. We prove, and confirm empirically, that this bound controls the solution quality end to end, recovering the classic $(1-1/\mathrm{e})$ guarantee up to client heterogeneity.
\item
\underline{Decentralization efficiency}: We avoid the need for the agents to download all agent aggregates and reach consensus among themselves. Because $\mathsf{SafeGM}$ decides in a random 2D projection and lifts the decision back, the contract can compute the decision easily from only two numbers per agent. Each client downloads only the few aggregates the contract selects. Our work makes classical depth-based robustness usable in high dimension for the first time.
\item
\underline{Empirical evidence}: On CIFAR10 and PathMNIST, with up to 30\% Byzantine clients and agents, $\mathsf{DeFSM}$ retains on average about 91\% of the quality of the geometric median, the most accurate robust aggregator in practice, at a trusted server, and never collapses, even under extreme heterogeneity. Decentralizing the geometric median, even in an idealized setting, reaches 96\% but costs six times more download per agent, or twice as much for clients; the mean and trimmed mean sometimes fall much lower.
\end{itemize}

Related work is reviewed in Section~\ref{sec:related} and FSM background in Section~\ref{sec:background}. $\mathsf{DeFSM}$ is detailed in Section~\ref{sec:solution}, analyzed in Section~\ref{sec:analysis}, and evaluated in Section~\ref{sec:eval}. Section~\ref{sec:conclusion} concludes the paper.

\section{Related Work}\label{sec:related}
Submodular Maximization (SM) is useful in machine learning, operations research, social networks, and network design, with applications such as active learning, feature selection, exemplar clustering, document summarization, object recognition, viral marketing, social welfare, sensor placement, and robotic data collection. 
 
\textbf{Conventional SM. } SM has been studied under many constraints, e.g., none, cardinality, knapsack, and non-linear constraints. The celebrated greedy algorithm of Nemhauser et al. \cite{10.1007/BF01588971} achieves the $(1-1/\mathrm{e})$ approximation ratio in polynomial time, the best possible unless P$=$NP \cite{feige1998threshold}. Distributed SM has also been explored \cite{pmlr-v37-barbosa15,10.5555/2946645.3053517, DBLP:conf/icml/Rafiey24,TranBigData2025,vendrell2024optimalitygapdecentralizedsubmodular}, mostly for computational efficiency under homogeneous conditions.

\textbf{Federated SM. } FSM research is rare \cite{DBLP:conf/icml/Rafiey24,10286024,TranBigData2025}. Rafiey \cite{DBLP:conf/icml/Rafiey24} proposed what appears to be the first FSM algorithm, a federated continuous greedy method, and proved its convergence. Wang et al. \cite{10286024} addressed client privacy in FSM. Tran et al. \cite{TranBigData2025} proposed an FSM design robust to practical client attacks, the only work on Byzantine-robust FSM to date. Client risks are widely studied in Federated Learning (FL), mostly as backdoor attacks \cite{nguyen2023backdoor}, but FSM faces attacks unseen in FL that require new defenses \cite{TranBigData2025}.

\textbf{Decentralized FSM. } When the server is untrustworthy or cannot be deployed, blockchain is a natural candidate to decentralize aggregation. No prior work uses blockchain for FSM, but its adoption in FL, though early, is rising \cite{9293091,10.1145/3579654.3579700,Madill2022ScaleSFLAS,TranBigData2024,10001370,10074277}. Existing designs build a blockchain network dedicated to FL or implement the server logic as a smart contract. Neither scales to the workloads of federated computation and storage. $\mathsf{DeFSM}$ instead outsources aggregation and storage to decentralized agents, with minimal contract overhead.

\textbf{Robust aggregation and vector consensus. } Byzantine-robust aggregation rules for FL, such as Krum \cite{10.5555/3294771.3294783}, the coordinate-wise trimmed mean \cite{Yin2018ByzantineRobustDL} and median \cite{10.1145/3709023.3737691}, the geometric median \cite{pillutla2022robust}, and median-of-means \cite{10.1007/978-3-031-25158-0_19}, work on full $d$-dimensional updates, rely on distances between updates or on coordinate-wise order statistics, and favor proximity to a majority; Krum and the trimmed mean also need the number of Byzantine updates as a parameter. Tailored attacks exploit this preference by keeping bad updates close to good ones while steering the aggregate \cite{DBLP:conf/icml/MhamdiGR18,10.1609/aaai.v37i7.26083,11023255}. In FSM, heterogeneous good clients' local solutions are not mutually close, so proximity is an even weaker sign of a good update \cite{TranBigData2025}. Byzantine vector consensus (BVC) \cite{10.1145/2484239.2484256,10.1145/2488608.2488657} instead requires the agreed vector to lie in the convex hull of the good agents' inputs, a distance-free guarantee, but is solvable only in very low dimension. Such guarantees rely on deep points, whose Tukey depth exceeds the number of bad inputs; they exist by the centerpoint theorem \cite{rado1947theorem} and are efficiently computable in the 2D plane \cite{chan2004optimal}. We make deep points usable for high-dimensional federated aggregation by deciding in 2D projections and bounding the error of lifting back. Unlike BVC, the agents in $\mathsf{DeFSM}$ never exchange messages to agree: the smart contract computes the deep point.

\section{Preliminaries} \label{sec:background}

 Consider a utility function $f: 2^E \to \mathbb{R}$ defined on all subsets of a finite ground set $E$. Function $f$ is \underline{submodular} iff for any subset $S \subset E$ and any elements $e_1, e_2 \in E$ we have 
\[
f(S \cup \{e_2\}) - f(S) \ge f(S \cup \{e_1, e_2\})-f(S \cup \{e_1\}).
\]
That is, $f$ exhibits diminishing returns: the marginal gain of adding an element decreases as the context set grows.  

\subsection{ {Centralized Submodular Maximization (SM)}}
 The goal of SM is to find a subset $S \subset E$ that maximizes $f(S)$ subject to a set constraint $\mathcal{I} \subset 2^E$: 
\[
\max_S \{ f(S): S \in \mathcal{I} \}.
\]
Ground set $E$ may be a collection of text documents for summarization or unlabeled images for annotation, while $f$ evaluates the quality of the chosen subset. Constraint $\mathcal{I}$ is often a \underline{matroid}; for example, the cardinality constraint $\mathcal{I} = \{S \subset E: |S| \le r\}$ is a matroid, whereas the knapsack constraint $\mathcal{I} = \{S \subset E \mid \sum_{s \in S} w_s \le b\}$ is not. Let $r \triangleq \max \{ |I| : I \in \mathcal{I}\}$ be the rank of constraint matroid $\mathcal{I}$. 

\textbf{Continuous greedy algorithm. } SM is strongly NP-hard, but it can be solved approximately up to an optimal approximation ratio of $(1-\frac{1}{{\mathrm{e}}})$ \cite{feige1998threshold}. This is achieved for the cardinality constraint by Nemhauser's seminal algorithm \cite{10.1007/BF01588971}. For arbitrary matroid constraints, the approach is to convert the discrete set function to a continuous version so that a gradient ascent method can be applied \cite{10.1145/1374376.1374389,10.1137/080733991}.
A subset $S \subset E$ is represented by its indicator vector $\mathbf{x}_S \in [0,1]^{|E|}$, with $\mathbf{x}[e]= 1 \Leftrightarrow e \in S$; conversely, an arbitrary vector $\mathbf{x}\in[0,1]^{|E|}$ corresponds to a random set $S$ containing each $e$ independently with probability $\mathbf{x}[e]$. The multilinear extension of $f$ is
\[
F(\mathbf{x}) \triangleq \mathbb{E}_{S \sim \mathbf{x}} [f(S)] = \sum_{S \subset E} f(S) \prod_{e \in S} \mathbf{x}[e]\prod_{e \not \in S} (1-\mathbf{x}[e]).
\]
Because $f(S)=F(\mathbf{x}_S)$, maximizing $f$ over sets becomes maximizing $F$ over vectors. The gradient of $F$ has coordinates $\partial F/\partial \mathbf{x}[e] = \mathbb{E}_{S \sim \mathbf{x}}[ f(S \cup \{e\}) - f(S \setminus \{e\})]$, which can be estimated by sampling $f$. Starting from $\mathbf{x}^{(0)} = \mathbf{0}$, the algorithm iterates, for $t=0,1,\dots$,
\begin{align*}
\mathbf{w}^{(t)} &= \arg \max_{\mathbf{w} \in \mathcal{P}} \langle \mathbf{w}, \nabla F(\mathbf{x}^{(t)}) \rangle,\qquad
\mathbf{x}^{(t+1)} = \mathbf{x}^{(t)} + \eta \mathbf{w}^{(t)},
\end{align*}
where $\eta \in (0,1)$ is the learning rate and $\mathcal{P}$ is the matroid polytope, the convex hull of the indicator vectors of the independent sets of $\mathcal{I}$. After $T=1/\eta$ steps, $\mathbf{x}$ is a convex combination of points of $\mathcal{P}$ and hence feasible. For the cardinality constraint $\mathcal{I} = \{ S \subset E: |S| \le r\}$, with $d\triangleq|E|$,
\[
\mathcal{P} = \Big\{\mathbf{x} \in [0, 1]^d: \textstyle\sum_{e \in E} \mathbf{x}[e] \le r \Big\}.
\]
Its vertices with exactly $r$ ones,
\[
\mathcal{V}_r\triangleq\Big\{\mathbf{w}\in\{0,1\}^d:\textstyle\sum_{e} \mathbf{w}[e]=r\Big\},
\]
are the indicator vectors of the $r$-subsets of $E$; for monotone objectives, the linear maximization above can always be solved by a vector of $\mathcal{V}_r$. The final subset can be taken as the $r$ largest coordinates of $\mathbf{x}$, or obtained by pipage rounding \cite{10.1137/080733991}, which preserves the approximation guarantee in expectation.

\subsection{ {Federated Submodular Maximization (FSM)}}
 Consider a standard federated setting with a server and $N$ clients $\mathcal{C}=\{1,\dots,N\}$, each identified with its index. Client $i$ has a local submodular function $f_i: 2^E \to \mathbb{R}$ to value a candidate subset $S$. The objective is to find a subset of $E$ under constraint $\mathcal{I}$ that maximizes the average of these local functions, i.e.,
 \[
   \max_{S \in \mathcal{I}} f(S) \triangleq \frac{1}{N} \sum_{i \in \mathcal{C}} f_i(S). 
 \]
Objective $f$ is submodular as an average of submodular functions. Ground set $E$ and constraint $\mathcal{I}$ are globally known, but $f_i$ is private to client $i$, and clients communicate only with the server. The continuous greedy algorithm for SM is extended for FSM as follows \cite{DBLP:conf/icml/Rafiey24}.

\textbf{Federated continuous greedy algorithm. } Let $F$ and $F_i$ be the multilinear extensions of $f$ and $f_i$, respectively. 
FSM runs $T$ ``aggregation rounds''. In round $t$, it keeps a global solution ($\mathbf{x}^{(t)}, \mathbf{w}^{(t)})$ at the server and a local one ($\mathbf{x}_i^{(t)}, \mathbf{w}^{(t)}_i$) at each client $i$, all zero at $t=0$. To begin a new round $t=1,2,...$, the server broadcasts the global gradient $\mathbf{w}^{(t-1)}$ to a subset $\mathcal{C}^{(t)} \subset \mathcal{C}$ of clients (typically a random subset of fixed size). Upon receipt, each client $i \in \mathcal{C}^{(t)}$ runs one greedy step to improve the solution using local continuous function $F_i$:
\begin{equation}
\mathbf{x}^{(t-1)} = \mathbf{x}^{(t-2)} + \eta \mathbf{w}^{(t-1)},\quad
\mathbf{w}_i^{(t)} = \arg \max_{\mathbf{w} \in \mathcal{P}} \langle \mathbf{w}, \nabla F_i(\mathbf{x}^{(t-1)}) \rangle,\quad
\mathbf{x}_i^{(t)} = \mathbf{x}^{(t-1)} + \eta \mathbf{w}_i^{(t)}.
\label{eq:fedCG_localsolution_update}
\end{equation}
Then client $i$ uploads its (polytope-projected) gradient vector $\mathbf{w}_i^{(t)}$ to the server. Having received these vectors from all clients, the server updates the global solution:
\begin{equation}
\mathbf{w}^{(t)} = \mathsf{mean} \big( \{\mathbf{w}_i^{(t)}: i \in \mathcal{C}^{(t)} \} \big),\qquad
\mathbf{x}^{(t)} = \mathbf{x}^{(t-1)} + \eta \mathbf{w}^{(t)}.
\label{eq:aggregate_mean}
\end{equation}
The new gradient aggregate $\mathbf{w}^{(t)}$ is broadcast to clients again in round $(t+1)$.

\section{{DeFSM}: Decentralized FSM with {SafeGM}}\label{sec:solution}
FSM is exposed to misbehavior by both the server and the clients. We propose $\mathsf{DeFSM}$, a server-less implementation of FSM in which the server is replaced with a lightweight smart contract and a pool of independent agents. The central question is how to compute the global aggregate of Eq.~\eqref{eq:aggregate_mean} in the presence of Byzantine clients and Byzantine agents, with a smart contract that can afford only light computation. The arithmetic mean is not robust, and robust aggregators work in $\mathbb{R}^d$, which the contract cannot afford. $\mathsf{DeFSM}$ therefore makes the robust decision in a 2D projection and lifts it back; safety, defined next, is what makes the lift trustworthy.

\subsection{Threat Model}\label{sec:threatmodel}
\textbf{Participants. } We call uncompromised (compromised) clients \emph{good} (\emph{bad}); likewise for agents. There are $M$ agents, indexed like the clients, with set $\mathcal{A}=\{1,\dots,M\}$. We mark good with $+$ and bad with $-$: $\Cg,\Cb\subseteq\mathcal{C}$ are the sets of good and bad clients and $N^+,N^-$ their numbers; likewise, $\Ag,\Ab\subseteq\mathcal{A}$ are the sets of good and bad agents and $M^+,M^-$ their numbers. The Byzantine client rate is $\beta_c=N^-/N$ and the Byzantine agent rate is $\beta_a=M^-/M$. In a given round, let $\mathbf{W}=\{\mathbf{w}_j\}_{j=1}^{N}$ be the submitted local updates, $\mathbf{w}_j$ being the update of client $j$; for brevity, we also write $\Cg$ and $\Cb$ for the sets of good and bad updates, $\{\mathbf{w}_j:j\in\Cg\}$ and $\{\mathbf{w}_j:j\in\Cb\}$. 

\textbf{Adversary. } The adversary knows all protocol logic and coordinates all bad clients and bad agents. Bad clients may observe the good updates before submitting their own (rushing adversary) and choose their updates arbitrarily; bad agents may deviate arbitrarily from the protocol: post any agent aggregate and any 2D point, submit invalid answers to the contract, or withhold data. The adversary cannot alter the code run by good participants or the smart contract, forge signatures, or break the hash function, and the local functions $f_i$ of good clients are private. The contract draws the agents of each round at random from a pool of registered participants with deposits, and so the adversary controls how many of its accounts are in the pool but not which are drawn.

\textbf{Validity filter. } Every good local update is the output of a linear maximization over $\mathcal{P}$, attained at a vertex; for the cardinality constraint with monotone objectives, it is the indicator of the $r$ elements with the largest marginal gains (ties broken by index), a vector in $\mathcal{V}_r$ (Section~\ref{sec:background}). Agents therefore discard any submitted vector outside $\mathcal{V}_r$, an $O(d)$ check, so that, without loss of generality, every bad update is also a binary vector with exactly $r$ ones: the adversary chooses only \emph{which} $r$ elements to select. For a general matroid, the filter accepts indicators of bases, checkable with an independence oracle.

\subsection{Safety as the Aggregation Target}\label{sec:safety}
\textbf{Safety target. } Deciding in a low-dimensional projection and lifting back loses information. Therefore, we need a guarantee that the lift preserves, at least approximately, whatever the adversary does. Distances are a poor basis, since attacks can shape them. We target an aggregate that is \emph{safe}: a convex combination of good updates only, i.e., a point of their convex hull. A robust aggregate may still give bad updates a positive share; a safe aggregate gives them none.

\begin{definition}[Safe vector, safety deficit]\label{def:safety}
A vector $\mathbf{w}$ is \emph{safe} if $\mathbf{w}\in\operatorname{conv}({\Cg})$. For $\mathbf{w}\in\mathcal{P}$, its \emph{safety deficit} is
\[
\sigma_{\Cg}(\mathbf{w}) \triangleq \min\big\{s\in[0,1] : \mathbf{w}\in(1-s)\operatorname{conv}({\Cg})+s\,\mathcal{P}\big\}.
\]
\end{definition}
The deficit is the smallest fraction of $\mathbf{w}$ that cannot be attributed to good clients; $\sigma_{\Cg}(\mathbf{w})=0$ iff $\mathbf{w}$ is safe, and $\sigma_{\Cg}\le 1$ always since $\operatorname{conv}({\Cg})\subseteq\mathcal{P}$. It does not depend on how $\mathbf{w}$ was computed. For example, the arithmetic mean of validity-filtered updates has $\sigma_{\Cg}\le\beta_c$, with equality when bad updates lie outside $\operatorname{conv}({\Cg})$; a selection rule such as Krum has $\sigma_{\Cg}\in\{0,1\}$ in each round depending on whether it selects a good update.

Section~\ref{sec:analysis} (Theorem~\ref{thm:safety-suffices}) will later show why this is the right target for FSM: for monotone objectives, an aggregate with deficit $\sigma$ can only \emph{dilute} the greedy progress by a factor $(1-\sigma)$, regardless of what the adversary does; a safe aggregate in every round recovers the $(1-1/\mathrm{e})$ guarantee up to client heterogeneity. 
Exact safety in $\mathbb{R}^d$, however, cannot be certified efficiently. The classical certificate is Tukey depth.

\begin{definition}[Tukey depth, deep point, Tukey median]\label{def:depth}
The \emph{Tukey depth} of a point $\mathbf{u}\in\mathbb{R}^k$ with respect to a finite multiset $\mathbf{U}\subset\mathbb{R}^k$ is the minimum number of points of $\mathbf{U}$ in a closed half-space containing $\mathbf{u}$:
\[
\operatorname{depth}(\mathbf{u};\mathbf{U}) \triangleq \min_{\mathbf{y}\in\mathbb{R}^k\setminus\{\mathbf{0}\}} \big|\{\mathbf{v}\in\mathbf{U}: \langle \mathbf{y},\mathbf{v}\rangle\ge\langle \mathbf{y},\mathbf{u}\rangle\}\big|.
\]
A point $\mathbf{u}$ is a \emph{deep point} if its depth exceeds the number of bad points in $\mathbf{U}$. A point of maximum depth is a \emph{Tukey median}.
\end{definition}

A deep point is always safe: otherwise, a closed half-space containing it would miss every good point and hold fewer points than its depth.
In $\mathbb{R}^d$, a deep point is guaranteed to exist only when $\beta_c<1/(d+1)$, which is vacuous when $d$ is the size of the ground set. We therefore \emph{target} safety through a tractable proxy that is exact in a 2D Gaussian random projection, and controls what is lost when lifting back to $\mathbb{R}^d$. We do not claim exact safety in $\mathbb{R}^d$.

\subsection{The $\mathsf{SafeGM}$ Aggregation Rule}\label{sec:psa}
\textbf{Notation.} A superscript $+$ marks a safe (or safety-targeting) vector, $*$ a vector of the global (agent-level) aggregation, and a hat a vector lifted from the projected space, hence an approximation; symbols without a hat are exact. For instance, $\mathbf{u}^+$ targets safety exactly in the projection, while its lift $\hat{\mathbf{w}}^+$ approximates a safe vector. 

Let $\Pi:\mathbb{R}^d\to\mathbb{R}^2$ be the round's 2D Gaussian random projection ($\Pi_{ij}\sim\mathcal{N}(0,1/2)$). We choose the plane for two reasons. First, a Tukey median of $n$ points is cheap to compute in 2D, taking $O(n\log n)$ expected time \cite{chan2004optimal}. Second, by the centerpoint theorem \cite{rado1947theorem}, the plane tolerates a Byzantine client rate up to $1/3$, higher than $1/(k+1)$ in $\mathbb{R}^k$ for $k>2$. A Byzantine threshold $1/3$ is the standard assumption of Byzantine fault-tolerant systems \cite{10.1145/357172.357176,10.5555/296806.296824}.

Our aggregation rule $\mathsf{SafeGM}$ is:
\begin{enumerate}
\item \emph{Filter, deduplicate, and project:} discard $\mathbf{w}_j\notin\mathcal{V}_r$, keep one copy of each distinct update, and apply the 2D projection $\mathbf{u}_j=\Pi\mathbf{w}_j$, $\mathbf{U}=\{\mathbf{u}_j\}$.
\item \emph{Depth test:} compute the geometric median $\mathbf{g}$ of $\mathbf{U}$. If $\operatorname{depth}(\mathbf{g};\mathbf{U})\ge\lceil n/3\rceil$, set $\mathbf{u}^+=\mathbf{g}$; otherwise let $\mathbf{u}^+$ be a Tukey median of $\mathbf{U}$.
\item \emph{Capped lifting:} solve the linear program
\begin{equation}
\min_{\boldsymbol{\lambda},\tau}\ \tau \quad \text{s.t.}\ \ \textstyle\sum_j\lambda_j\mathbf{u}_j=\mathbf{u}^+,\quad \sum_j\lambda_j=1,\quad 0\le\lambda_j\le\tau, \label{eq:cappedLP}
\end{equation}
and output $\hat{\mathbf{w}}^+=\sum_j\lambda_j\mathbf{w}_j$. 
\end{enumerate}

Deduplication in Step~1 gives every distinct update one vote. Bad clients that coordinate on the same update, as in the Include attack \cite{TranBigData2025}, would otherwise stack their copies on a single projected point, which then gains depth and coefficient in proportion to the number of copies. Heterogeneous good clients rarely submit identical updates, and removing a copy of a good update cannot make the aggregate unsafe. Hereafter, ${\Cg}$ and ${\Cb}$ denote the \emph{distinct} good and bad updates, $n=|{\Cg}|+|{\Cb}|$, and an update identical to a good one counts as good.

In Step~2, we use the geometric median first because it is the most accurate of the common robust aggregators in practice \cite{pillutla2022robust}, and in the plane it is almost always deep: in our experiments its depth reached $\lceil n/3\rceil$ in about 96\% of aggregations. The Tukey median is only a fallback that ensures depth at least $\lceil n/3\rceil$ in every case (see Lemma~\ref{lem:capped}(i) below), which is all the analysis needs. The rule never uses the number of bad updates. Whenever that number is below the depth of $\mathbf{u}^+$, $\mathbf{u}^+$ is a deep point and hence safe, i.e., $\mathbf{u}^+\in\Pi\operatorname{conv}({\Cg})$. Lemma~\ref{lem:capped} below bounds the share of bad updates in every case.

\begin{lemma}[Depth and capped lifting]\label{lem:capped}
Let $D^+\triangleq\operatorname{depth}(\mathbf{u}^+;\mathbf{U})$. Then:
(i) $D^+\ge\lceil n/3\rceil$;
(ii) the optimal value of the capped linear program \eqref{eq:cappedLP} is at most $1/D^+$, hence every coefficient $\lambda_j\le1/D^+$;
and
(iii) the safety deficit of the output $\hat{\mathbf{w}}^+$ satisfies
\[
\sigma_{\Cg}(\hat{\mathbf{w}}^+)\ \le\ \lambda({\Cb})\ \le\ \frac{|{\Cb}|}{D^+},
\]
where $\lambda({\Cb})\triangleq\sum_{j\in {\Cb}}\lambda_j$ is the share of bad updates.

\end{lemma}

\begin{proof}
(i) If $\mathbf{u}^+=\mathbf{g}$, this is the test in Step~2. Otherwise $\mathbf{u}^+$ is a Tukey median, whose depth is at least $\lceil n/3\rceil$ by the centerpoint theorem in the plane \cite{rado1947theorem}.

(ii) Let $\mathcal{Q}=\{\sum_j\lambda_j\mathbf{u}_j:\boldsymbol{\lambda}\in\Delta_n,\ \lambda_j\le1/D^+\}$, a compact convex set whose support function is $h_{\mathcal{Q}}(\mathbf{y})=\frac1{D^+}\sum_{\text{$D^+$ largest}}\langle\mathbf{y},\mathbf{u}_j\rangle$ (fill the cap greedily). By the definition of depth, for every $\mathbf{y}\neq\mathbf{0}$ at least $D^+$ points satisfy $\langle\mathbf{y},\mathbf{u}_j\rangle\ge\langle\mathbf{y},\mathbf{u}^+\rangle$, and so the average of the $D^+$ largest values is at least $\langle\mathbf{y},\mathbf{u}^+\rangle$. Hence $\langle\mathbf{y},\mathbf{u}^+\rangle\le h_{\mathcal{Q}}(\mathbf{y})$ for all $\mathbf{y}$ (trivially for $\mathbf{y}=\mathbf{0}$), and so $\mathbf{u}^+\in\mathcal{Q}$. The corresponding $\boldsymbol{\lambda}$, with $\tau=1/D^+$, is feasible for \eqref{eq:cappedLP}. The optimal value $\tau^*$ is therefore at most $1/D^+$, and the returned coefficients satisfy $\lambda_j\le\tau^*\le1/D^+$.

(iii) Summing (ii) over the bad updates gives $\lambda({\Cb})\le|{\Cb}|/D^+$. If $\lambda({\Cb})=0$, then $\hat{\mathbf{w}}^+\in\operatorname{conv}({\Cg})$ and $\sigma_{\Cg}(\hat{\mathbf{w}}^+)=0$. If $0<\lambda({\Cb})<1$, write $\hat{\mathbf{w}}^+=(1-\lambda({\Cb}))\,\mathbf{g}'+\lambda({\Cb})\,\mathbf{p}$ with $\mathbf{g}'=\sum_{j\in {\Cg}}\lambda_j\mathbf{w}_j/(1-\lambda({\Cb}))\in\operatorname{conv}({\Cg})$ and $\mathbf{p}=\sum_{j\in {\Cb}}\lambda_j\mathbf{w}_j/\lambda({\Cb})\in\mathcal{P}$, since bad updates pass the validity filter; hence $\sigma_{\Cg}(\hat{\mathbf{w}}^+)\le\lambda({\Cb})$. The case $\lambda({\Cb})=1$ is trivial.
\end{proof}
\textbf{Remark. } The bound on the share of bad updates holds in $\mathbb{R}^d$: whatever the bad updates are and wherever their projections lie, together they carry at most a $|{\Cb}|/D^+$ share of $\hat{\mathbf{w}}^+$. By Lemma~\ref{lem:capped}(i), this fraction is at most three times the fraction of bad updates. The proof above uses only the depth of $\mathbf{u}^+$, and so the same bound, with $D$ in place of $D^+$, holds for any point of depth $D\ge1$ and any coefficients $\lambda_j\le1/D$ that reproduce it. In Section~\ref{sec:analysis}, we turn this bound into a guarantee on solution quality (Theorem~\ref{thm:safety-suffices}) and apply it at the agent level (Theorem~\ref{thm:decentral}).

Unlike Krum- and trimming-based rules, which require knowing a priori how many updates to discard, $\mathsf{SafeGM}$ requires no knowledge of the Byzantine client rate: its tolerance adapts to the data through the depth $D^+$, which is often well above the test threshold $\lceil n/3\rceil$.
In the global update, $\hat{\mathbf{w}}^+$ replaces the mean,
\begin{equation}
\mathbf{w}^{(t)} = \mathsf{SafeGM}\big(\{\mathbf{w}_i^{(t)} \mid i \in \mathcal{C}^{(t)}\}\big),\qquad
\mathbf{x}^{(t)} = \mathbf{x}^{(t-1)} + \eta \mathbf{w}^{(t)}. \label{eq:tverberglogic}
\end{equation}

\subsection{Decentralization}\label{sec:decentral}

We now decentralize the safe aggregation. 
$\mathsf{DeFSM}$ splits the work between the smart contract and the off-chain agents so that each party does only what it can do cheaply. Agents aggregate random samples of client updates, and the contract makes the robust decision on 2D projections of the agents' results. This split is possible because $\mathsf{SafeGM}$ decides in the plane: the contract only needs $2M$ numbers (the 2D coordinates of each agent's projection), a geometric median or Tukey median of $M$ points in the plane, and a small linear program. In contrast, an aggregation rule that decides in $\mathbb{R}^d$ would need the contract to store $Md$ numbers and run the rule on them. Its agents would have to compute it off-chain and vote on the result instead.

\textbf{Agents. } Being an agent is a role, not a separate kind of machine. Anyone can register for it by placing a deposit with the contract, including clients. In each round, the contract draws $M$ agents at random from the registered pool. This draw takes place after the client deadline, and as the set of agents differs from round to round, an adversary cannot choose which of its accounts are drawn and no agent knows its sample in advance.

\subsubsection{Protocol}\label{sec:twolevel}
\begin{figure}[tbp]
    \centering
    \includegraphics[width=\textwidth]{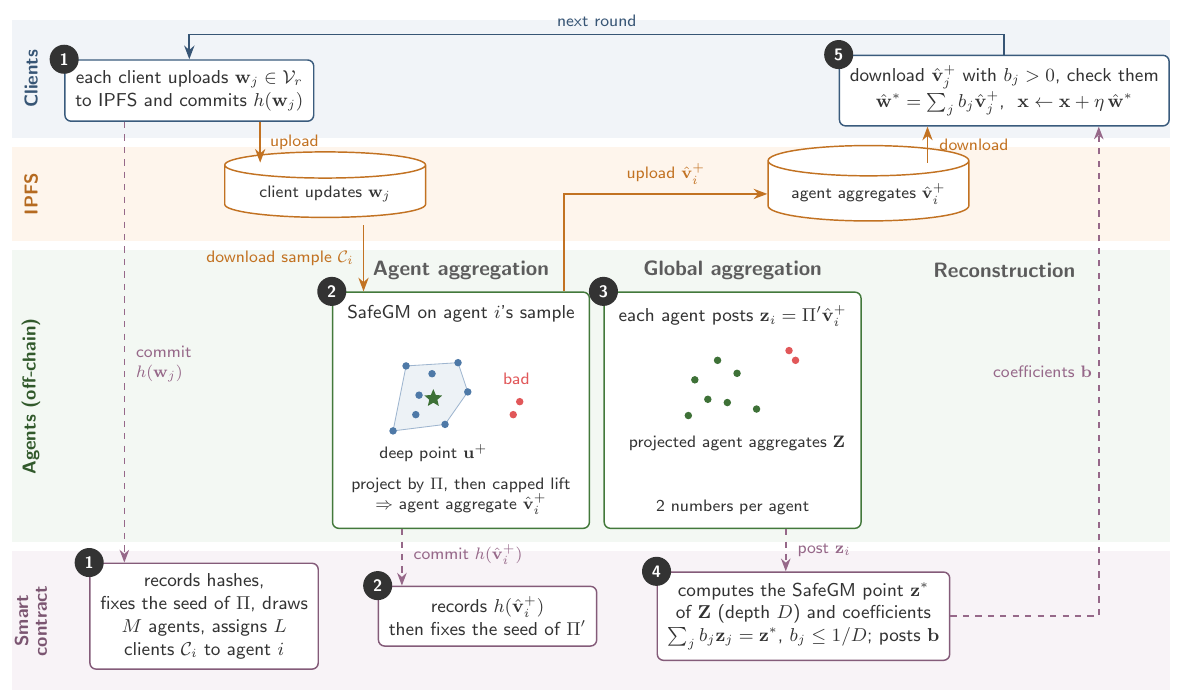}
    \caption{One round of $\mathsf{DeFSM}$. Solid arrows are off-chain data transfers (orange: to and from IPFS); dashed arrows are on-chain posts. Numbers match the steps of the protocol.}
    \label{fig:protocol}
\end{figure}

Each round has three main tasks. 
In \emph{agent aggregation}, each agent runs $\mathsf{SafeGM}$ on a random sample of $L\ll N$ client updates. In \emph{global aggregation}, the contract computes the $\mathsf{SafeGM}$ point of the agents' 2D projections and coefficients that reproduce it; no agreement among agents is needed. In \emph{reconstruction}, each client computes the global aggregate as the weighted sum of the agent aggregates with positive coefficient.
There are two projection matrices, $\Pi$ for projecting client updates and $\Pi'$ for projecting agent aggregates. 
Fig.~\ref{fig:protocol} illustrates the steps. In detail, a round proceeds as follows.
\begin{enumerate}
\item \underline{Client commit.} Each client uploads $\mathbf{w}_j\in\mathcal{V}_r$ to IPFS and commits $h(\mathbf{w}_j)$ to the contract. The transaction closing the client deadline fixes the seed of the client-level projection $\Pi$. The contract then draws $M$ random agents and assigns to each agent $i$ a uniformly random sample $\mathcal{C}_i\subseteq\mathcal{C}$ of $L$ clients, drawn independently across agents.
\item \underline{Agent aggregation.} Agent $i$ downloads its assigned updates $\mathbf{W}_i=\{\mathbf{w}_{il}\}_{l=1}^L$ from IPFS, verifies their hashes and the validity filter, and runs $\mathsf{SafeGM}$ with $\Pi$, obtaining the agent aggregate $\hat{\mathbf{v}}^+_i=\sum_l a_{il}\mathbf{w}_{il}\in\mathcal{P}$. It uploads $\hat{\mathbf{v}}^+_i$ to IPFS and commits $h(\hat{\mathbf{v}}^+_i)$ to the contract.
\item \underline{Projection.} The transaction closing the agent deadline fixes the seed of the agent-level projection $\Pi'$. Each agent $i$ posts $\mathbf{z}_i=\Pi'\hat{\mathbf{v}}^+_i$ to the contract. The contract thus keeps two numbers (the 2D coordinates) per distinct commitment, and so agents with identical aggregates count once. Let $\mathbf{Z}=\{\mathbf{z}_j\}$ be the kept points.
\item \underline{Global aggregation.} The contract applies the 2D aggregate of $\mathsf{SafeGM}$ to $\mathbf{Z}$: the 2D aggregate $\mathbf{z}^*$ is the geometric median if its depth is at least $\lceil|\mathbf{Z}|/3\rceil$, or a Tukey median otherwise. It then computes convex coefficients $\mathbf{b}$ of $\mathbf{z}^*$ that solve \eqref{eq:cappedLP} with the cap $\tau=1/D$, 
where $D \triangleq \operatorname{depth}(\mathbf{z}^*;\mathbf{Z})$, using a basic solution of the linear program found with a fixed pivoting rule. 
This solution has at most three coefficients strictly between $0$ and $1/D$, one per equality constraint. Therefore, at most $D+2$ coefficients are positive, and so each client downloads at most $D+2$ of the $M$ agent aggregates, saving communication. The contract posts $\mathbf{b}$.

\item \underline{Reconstruction.} Each client downloads the agent aggregates $\{\hat{\mathbf{v}}^+_j \mid b_j>0\}$, verifies their hashes, and computes the global aggregate
\begin{align}
\hat{\mathbf{w}}^*=\sum_j b_j\,\hat{\mathbf{v}}^+_j=\sum_j b_j\sum_l a_{jl}\,\mathbf{w}_{jl},\label{eq:globalupdatex}
\end{align}
and the global step $\mathbf{x}\leftarrow\mathbf{x}+\eta\hat{\mathbf{w}}^*$. The coefficients are on-chain and the sum is computed in fixed-point arithmetic, and so all good clients obtain the same $\hat{\mathbf{w}}^*$ and the same $\mathbf{x}$. A client also checks that $\Pi'\hat{\mathbf{v}}^+_j=\mathbf{z}_j$ and $\hat{\mathbf{v}}^+_j\in\mathcal{P}$ for each downloaded aggregate. If a check fails, it posts the aggregate as a fraud proof; the contract removes the agent's point, slashes its deposit, and the round repeats Step~4.
\end{enumerate}

No step of the basic protocol requires the agents to communicate, reach consensus, or vote: the contract makes the robust decision at the agent level, on public data.

\textbf{Bandwidth-limited clients. } In Step~5, a client downloads about $D$ agent aggregates. If clients cannot afford this, one agent can compute $\hat{\mathbf{w}}^*$, upload it to IPFS, and commit its hash; the agents then vote on the hash, and each client downloads a single vector. This vote certifies only the arithmetic of a weighted sum whose coefficients are already fixed on-chain; it does not affect the robust decision. We use client-side reconstruction, which needs no vote.

\subsubsection{Complexity}

Consider a round, with sparse binary updates of $r$ ones.
For \emph{agents}, each agent downloads $L$ client updates and spends $O(Ld)$ time for the projection, $O(L\log L)$ time for the depth test, and $O(L)$ time for the capped lift \cite{megiddo1984linear}; hence $O(Ld)$ time in total, since $d\ge\log L$. Over all $M$ agents, this is $O(MLd)$ time and $ML$ downloads. All these costs are independent of $N$.
The \emph{contract} spends $O(M\log M)$ time, or $O(M^3\log M)$ in the rare fallback to a Tukey median, plus $O(M)$ time for its linear program; storage is $O(N+M)$, independent of $d$.
For the \emph{clients}, each downloads at most $D+2$ agent aggregates, each with at most $Lr$ nonzero entries, that is, $O(DLr)$ entries, and computes their weighted sum in the same time, independent of $d$ and $N$.
For example, in our experiments ($M=L=20$, $r=10$, $d=200$), a client downloads on average 9.3 agent aggregates with 487 nonzero entries per round in total, about 4~KB, less than a typical webpage.

\subsubsection{Unpredictable projections}\label{sec:unpredictable}
Each round uses two fresh projections: $\Pi$ for $\mathsf{SafeGM}$ (Step~1) and $\Pi'$ for the agent level (Step~3). Every party regenerates both from public 32-byte seeds fixed by the smart contract, from the chain's built-in randomness or a verifiable random function, so they cost no extra communication. Each seed is fixed only after the inputs it projects have been committed. Hence bad clients choose their updates without knowing $\Pi$, so they cannot place their projected points where they would gain the largest share, such as at the Tukey median; likewise, bad agents commit their aggregates without knowing $\Pi'$. The consistency check in Step~5 keeps them from moving their 2D points after $\Pi'$ is known. A projection reused across rounds would be known in advance.

A bad client also has little control over where its update lands in the plane. For the matrix $\mathbf{W}=[\mathbf{w}_1,\dots,\mathbf{w}_n]$ of distinct updates, the rows of $\mathbf{U}=\Pi\mathbf{W}$ are i.i.d. $\mathcal{N}(\mathbf{0},\mathbf{G}/2)$ with Gram matrix $\mathbf{G}=\mathbf{W}^\top\mathbf{W}$, and every entry of $\mathbf{G}$ lies in $[0,r]$ after the validity filter. Hence a bad client affects the projected points only through how many elements its update shares with the others, and cannot aim its update at the deep region. The same holds for $\Pi'$ and the agents' aggregates.

\section{Analysis}\label{sec:analysis}
We first show that the safety deficit controls the solution: the quality of the federated continuous greedy solution degrades gracefully with the deficit of the aggregates, independently of the attack (Section~\ref{sec:suffices}), so a bound on the deficit is all we need. We then bound the deficit of $\mathsf{SafeGM}$'s aggregate in $\mathsf{DeFSM}$ by counts and depths, and combine the two results into a convergence guarantee (Section~\ref{sec:decguarantee}).

\subsection{Safety Suffices for Monotone FSM}\label{sec:suffices}
Throughout, each good client's $f_i$ is monotone submodular with $f_i(\emptyset)=0$; $F_{\Cg}\triangleq\frac{1}{N^+}\sum_{i\in {\Cg}}F_i$ is the multilinear extension of the good clients' objective, $\mathrm{OPT}=\max_{S\in\mathcal{I}}\frac{1}{N^+}\sum_{i\in {\Cg}}f_i(S)$, and $m\triangleq\max_{i\in {\Cg},e\in E}f_i(\{e\})$. The algorithm runs $T=1/\eta$ rounds from $\mathbf{x}^{(0)}=\mathbf{0}$, with $\mathbf{x}^{(t)}=\mathbf{x}^{(t-1)}+\eta\mathbf{w}^{(t)}$. In round $t$, each good client $j$ returns $\mathbf{w}_j\in\arg\max_{\mathbf{w}\in\mathcal{P}}\langle\mathbf{w},\nabla F_j(\mathbf{x}^{(t-1)})\rangle$.

Like most analyses of federated optimization under non-iid data, we bound how far a client's gradient can be from the average \cite{DBLP:conf/mlsys/LiSZSTS20,DBLP:conf/iclr/LiHYWZ20}.
\begin{assumption}[Bounded heterogeneity]\label{ass:het}
There is $\zeta\ge0$ such that $\|\nabla F_i(\mathbf{x})-\nabla F_{\Cg}(\mathbf{x})\|_\infty\le\zeta$ for all $\mathbf{x}\in\mathcal{P}$ and $i\in {\Cg}$. (If clients use sampled gradient estimates, $\zeta$ also absorbs the estimation error.)
\end{assumption}

The next theorem shows that the adversary affects the solution only through the safety deficit of the aggregates.
\begin{theorem}[Safety suffices]\label{thm:safety-suffices}
Suppose every aggregate satisfies $\mathbf{w}^{(t)}\in\mathcal{P}$ and has safety deficit $\sigma_{{\Cg}_t}(\mathbf{w}^{(t)})\le\sigma_t$ for some $\sigma_t\in[0,1]$, where ${\Cg}_t$ is the set of good updates of round $t$, and let $\bar\sigma=\frac1T\sum_t\sigma_t$ be the average deficit bound. Then, for any behavior of the bad clients, the multilinear extension $F_{\Cg}$ of the good clients' objective satisfies
\[
F_{\Cg}(\mathbf{x}^{(T)})\ \ge\ \big(1-\mathrm{e}^{-(1-\bar\sigma)}\big)\,\mathrm{OPT}\;-\;2r\zeta\;-\;\frac{\eta\, m\, r^2}{2}.
\]
\end{theorem}
\begin{proof}
\emph{Feasibility.} $\mathbf{x}^{(t)}=\eta\sum_{s\le t}\mathbf{w}^{(s)}$ is $t\eta\le1$ times an average of points of $\mathcal{P}$, and $\mathbf{0}\in\mathcal{P}$, so $\mathbf{x}^{(t)}\in\mathcal{P}$.

\emph{Good directions are good ascent directions.} Let $\mathbf{x}=\mathbf{x}^{(t-1)}$ and $O$ an optimal set. For $j\in {\Cg}_t$, since $\mathbf{1}_O\in\mathcal{P}$ and $\|\mathbf{w}\|_1\le r$ on $\mathcal{P}$,
\begin{align*}
\langle\mathbf{w}_j,\nabla F_{\Cg}(\mathbf{x})\rangle
&\ge\langle\mathbf{w}_j,\nabla F_j(\mathbf{x})\rangle-r\zeta
\ge\langle\mathbf{1}_O,\nabla F_j(\mathbf{x})\rangle-r\zeta\\
&\ge\langle\mathbf{1}_O,\nabla F_{\Cg}(\mathbf{x})\rangle-2r\zeta
\ge\mathrm{OPT}-F_{\Cg}(\mathbf{x})-2r\zeta,
\end{align*}
where the last step is the standard property of monotone submodular multilinear extensions \cite{10.1137/080733991}: $F_{\Cg}(\mathbf{x}\vee\mathbf{1}_O)-F_{\Cg}(\mathbf{x})\le\langle\mathbf{1}_O,\nabla F_{\Cg}(\mathbf{x})\rangle$ by concavity along nonnegative directions and $\nabla F_{\Cg}\ge\mathbf{0}$, and $F_{\Cg}(\mathbf{x}\vee\mathbf{1}_O)\ge\mathrm{OPT}$ by monotonicity. By linearity the same bound holds for every $\mathbf{g}\in\operatorname{conv}({\Cg}_t)$.

\emph{Unsafe mass can only dilute.} Write $\mathbf{w}^{(t)}=(1-\sigma_t)\mathbf{g}+\sigma_t\mathbf{p}$ with $\mathbf{g}\in\operatorname{conv}({\Cg}_t)$, $\mathbf{p}\in\mathcal{P}$ (possible since $(1-s)\operatorname{conv}({\Cg}_t)+s\mathcal{P}$ grows with $s$ when $\operatorname{conv}({\Cg}_t)\subseteq\mathcal{P}$). Monotonicity gives $\nabla F_{\Cg}\ge\mathbf{0}$, and $\mathbf{p}\ge\mathbf{0}$, so $\langle\mathbf{p},\nabla F_{\Cg}(\mathbf{x})\rangle\ge0$ and
\[
\langle\mathbf{w}^{(t)},\nabla F_{\Cg}(\mathbf{x})\rangle\ge(1-\sigma_t)\big(\mathrm{OPT}-F_{\Cg}(\mathbf{x})\big)-2r\zeta .
\]
\emph{Smoothness.} For the multilinear extension of a monotone submodular function, $\partial^2F_{\Cg}/\partial x_e^2=0$ and $-m\le\partial^2F_{\Cg}/\partial x_e\partial x_{e'}\le0$, so, using $\|\mathbf{w}\|_1\le r$,
\[
F_{\Cg}(\mathbf{x}+\eta\mathbf{w})\ge F_{\Cg}(\mathbf{x})+\eta\langle\mathbf{w},\nabla F_{\Cg}(\mathbf{x})\rangle-\tfrac{\eta^2mr^2}{2}.
\]

\emph{Recursion.} With $\gamma_t=\mathrm{OPT}-F_{\Cg}(\mathbf{x}^{(t)})$, the last two displays give
\[
\gamma_t\le\big(1-\eta(1-\sigma_t)\big)\gamma_{t-1}+2r\zeta\eta+\tfrac{\eta^2mr^2}{2}.
\] Unrolling from $\gamma_0=\mathrm{OPT}$ and using $1-z\le\mathrm{e}^{-z}$ and $T\eta=1$,
\[
\gamma_T\le\mathrm{e}^{-\eta\sum_t(1-\sigma_t)}\mathrm{OPT}+2r\zeta+\tfrac{\eta mr^2}{2}
=\mathrm{e}^{-(1-\bar\sigma)}\mathrm{OPT}+2r\zeta+\tfrac{\eta mr^2}{2}. \qedhere
\]
\end{proof}

Theorem~\ref{thm:safety-suffices} formalizes our design philosophy. The bound depends on the adversary only through the deficit, not through any notion of distance, and the validity filter ensures that unsafe mass slows progress but never reverses it. For example, the filtered mean has $\bar\sigma\le\beta_c$, while a rule that selects a bad update has $\sigma_t=1$ and makes no guaranteed progress in that round, which is how distance-based selection collapses under targeted attacks. $\mathsf{SafeGM}$ aims at $\sigma_t\to0$. Since $1-\mathrm{e}^{-z}$ is concave, $1-\mathrm{e}^{-(1-\bar\sigma)}\ge(1-\bar\sigma)(1-1/\mathrm{e})$: an average deficit $\bar\sigma$ costs at most a factor $(1-\bar\sigma)$, and a safe aggregate in every round ($\bar\sigma=0$) recovers the centralized $(1-1/\mathrm{e})$ guarantee up to heterogeneity and discretization, for any adversary. The guarantee carries over to the final set: pipage rounding \cite{10.1137/080733991} turns $\mathbf{x}^{(T)}\in\mathcal{P}$ into a set $S$ with $|S|\le r$ and $\mathbb{E}[f_{\Cg}(S)]\ge F_{\Cg}(\mathbf{x}^{(T)})$.

\textbf{Remark (why FSM, not federated learning). } Theorem~\ref{thm:safety-suffices} rests on structure that FSM has and federated learning (FL) lacks. First, monotonicity makes every coordinate of the gradient nonnegative, and every update lies in the nonnegative polytope $\mathcal{P}$, so the unsafe part of an aggregate can only dilute progress, never reverse it. In FL, gradients take both signs, so even a small unsafe share can push the model in a harmful direction, and a bounded deficit no longer bounds the damage. Second, the validity filter checks that each update is a vertex of $\mathcal{P}$; no comparable check exists for a real-valued gradient. Third, FSM updates are sparse binary vectors with $r$ ones, so a client can afford to download the $D+2$ agent aggregates it needs; dense FL models with millions of parameters would multiply this cost. A count-based bound is thus a natural fit for FSM, and its extension to FL would require new ideas.

In practice, the final set is often taken deterministically as the $r$ largest coordinates of $\mathbf{x}^{(T)}$ (top-$r$ rounding), which is not lossless. Coordinated attacks such as Include \cite{TranBigData2025} exploit exactly this step: bad clients that all select the same $r$ elements concentrate their mass on few coordinates, while non-iid good clients spread theirs. Theorem~\ref{thm:safety-suffices} does not cover top-$r$ rounding, but the deficit still limits such attacks. Call an element \emph{unsupported} if no good client selects it in any round; the next proposition bounds, by the deficit, how much mass such elements can receive.
\begin{proposition}[Top-$r$ rounding]\label{prop:topr}
Under the conditions of Theorem~\ref{thm:safety-suffices}, every unsupported element $e$ satisfies $x^{(T)}[e]\le\bar\sigma$, and the supported elements together carry mass at least $(1-\bar\sigma)r$.
\end{proposition}
\begin{proof}
Write $\mathbf{w}^{(t)}=(1-\sigma_t)\mathbf{g}+\sigma_t\mathbf{p}$ with $\mathbf{g}\in\operatorname{conv}({\Cg}_t)$, $\mathbf{p}\in\mathcal{P}$. For unsupported $e$, every good update has $w_j[e]=0$, so $g[e]=0$ and $w^{(t)}[e]=\sigma_tp[e]\le\sigma_t$; summing, $x^{(T)}[e]\le\eta\sum_t\sigma_t=\bar\sigma$. Every good update is in $\mathcal{V}_r$ and has its $r$ ones on supported elements, so $\mathbf{g}$ puts mass $r$ on them, and the supported mass of $\mathbf{x}^{(T)}$ is at least $\eta\sum_t(1-\sigma_t)r=(1-\bar\sigma)r$.
\end{proof}
Hence, if at least $r$ supported elements have $x^{(T)}[e]>\bar\sigma$, the top-$r$ set contains no unsupported element; with safe aggregates, unsupported elements receive no mass and are never selected. For the filtered mean, $\bar\sigma$ can reach $\beta_c$: identical Include updates give their target coordinates mass $\beta_c$, which outranks supported elements chosen by fewer than a $\beta_c/(1-\beta_c)$ fraction of good clients in every round. Lowering $\bar\sigma$ below $\beta_c$ is therefore what protects the final ranking.

\subsection{Guarantee of the Decentralized Aggregate}\label{sec:decguarantee}
By Theorem~\ref{thm:safety-suffices}, it remains to bound the deficit of the global aggregate $\hat{\mathbf{w}}^*$ that the clients compute in Step~5 of the protocol. For agent $j$, let $\mathcal{C}_j\subseteq\mathcal{C}$ be its sample, with good and bad parts ${\Cg}_j$ and ${\Cb}_j$, $N_j^-$ the number of distinct bad updates in it, and $D_j^+$ the depth of the point chosen by $\mathsf{SafeGM}$ on its sample. Call a good agent \emph{in-tolerance} if $N_j^-<D_j^+$. Let ${\tilde{\mathcal{A}}^{+}}\subseteq {\Ag}$ be the in-tolerance good agents and ${\tilde{\mathcal{A}}^{-}}$ the remaining agents: the bad agents ${\Ab}$ and the out-of-tolerance good agents, which the analysis treats as bad; $\tilde M^+$ and $\tilde M^-$ denote the numbers of agents in these two sets. The next theorem bounds the deficit by counts and depths only. It needs no assumption on the network, because the agents do not need to agree with each other.

\begin{theorem}[Decentralized safety bound]\label{thm:decentral}
Let $\mathbf{z}^*$, $D$, and $\mathbf{b}$ be the point, depth, and coefficients computed by the smart contract. Then $\hat{\mathbf{w}}^*\in\mathcal{P}$ and
\[
\sigma_{\Cg}(\hat{\mathbf{w}}^*)\ \le\ \max_{j\in {\tilde{\mathcal{A}}^{+}}}\frac{N_j^-}{D_j^+}\;+\;\frac{\tilde M^-}{D}.
\]
\end{theorem}
\begin{proof}
Each agent aggregate that passes the checks of Step~5 lies in $\mathcal{P}$, and $\mathbf{b}$ lies in the simplex, so $\hat{\mathbf{w}}^*\in\mathcal{P}$. Each kept point $\mathbf{z}_j$ corresponds to one distinct agent aggregate $\hat{\mathbf{v}}_j$. Write $\hat{\mathbf{v}}_j=(1-\sigma_j)\mathbf{g}_j+\sigma_j\mathbf{p}_j$ with $\mathbf{g}_j\in\operatorname{conv}({\Cg})$ and $\mathbf{p}_j\in\mathcal{P}$. If an in-tolerance good agent produced $\hat{\mathbf{v}}_j$, then $\sigma_j\le\max_{i\in{\tilde{\mathcal{A}}^{+}}}N_i^-/D_i^+$ by Lemma~\ref{lem:capped}; otherwise $\sigma_j\le1$, and at most $\tilde M^-$ distinct aggregates are of this kind. The contract's coefficients satisfy $b_j\le1/D$, so these aggregates carry a total share at most $\tilde M^-/D$. Mixing the decompositions with coefficients $b_j$ gives $\sigma_{\Cg}(\hat{\mathbf{w}}^*)\le\sum_jb_j\sigma_j\le\max_{i\in{\tilde{\mathcal{A}}^{+}}}N_i^-/D_i^+ + \tilde M^-/D$.
\end{proof}

Theorem~\ref{thm:decentral} applies Lemma~\ref{lem:capped} at both levels: the first term bounds the lift from client updates to agent aggregates, and the second bounds the lift from the contract's 2D point to $\mathbb{R}^d$. The bound counts \emph{distinct} vectors at both levels, so copies of a bad update, or of a bad agent aggregate, add nothing. Bad agents enter only through their number, whatever they post. By Lemma~\ref{lem:capped}(i) applied to $\mathbf{Z}$, $D\ge\lceil|\mathbf{Z}|/3\rceil$, and so the agent-level term is at most three times $\tilde M^-/|\mathbf{Z}|$, and it is smaller when the depth of the contract's point is larger. Neither the clients, the agents nor the contract needs to know the Byzantine client or agent rate.

\textbf{Remark (role of unpredictable projections). } Lemma~\ref{lem:capped} and Theorem~\ref{thm:decentral} hold even if the adversary knows the projections in advance. Fresh, unpredictable projections (Section~\ref{sec:unpredictable}) only make their worst case harder to reach.

\textbf{Remark (why a deep point at the agent level). } The mean of the distinct agent aggregates has a smaller worst-case bound, $\tilde M^-/|\mathbf{Z}|$ instead of $\tilde M^-/D$. We use the capped deep point for two reasons. First, bad aggregates far from the deep region receive much smaller coefficients than the cap allows: at $\beta_c=\beta_a=0.3$, replacing it with the mean raised the realized deficit from 0.30 to 0.42 and lowered the quality from 0.79 to 0.68 on average over both datasets (Section~\ref{sec:eval}). Second, the coefficients are sparse: at most $D+2$ agent aggregates are needed instead of all $|\mathbf{Z}|$, which halves the client download in our experiments.

\begin{wraptable}{r}{0.3\textwidth}
\vspace{-1.2em}\centering\footnotesize
\caption{Worst-case probability $\rho_L$ that an agent's sample exceeds the planar tolerance.}
\label{tab:pl}
\begin{tabular}{cccc}
\toprule
$N$ & $\beta_c$ & $L$ & $\rho_L$\\
\midrule
60 & 1/6 & 20 & 0.012\\
60 & 1/3 & 20 & 0.534\\
200 & 1/6 & 30 & 0.011\\
200 & 1/4 & 40 & 0.079\\
\bottomrule
\end{tabular}
\vspace{-1em}
\end{wraptable}
\textbf{How likely is the tolerance condition? } Suppose the good updates within a sample are distinct, as is the case for heterogeneous clients. Then $|{\Cb}_j|<L/3$ implies $N_j^-<D_j^+$: the deduplicated sample has $|{\Cg}_j|>2L/3>2N_j^-$ good updates, so $N_j^-<(|{\Cg}_j|+N_j^-)/3\le D_j^+$ by the centerpoint theorem. The contract samples agents independently, so the number $M^+-\tilde M^+$ of out-of-tolerance good agents is dominated by $\mathrm{Bin}(M^+,\,\rho_L)$, where
\[
\rho_L=\Pr\big[\mathrm{Hyp}(N,N^-,L)\ge\lceil L/3\rceil\big]
\le\mathrm{e}^{-2L\left(\frac{1}{3}-\beta_c\right)^2}\qquad\big(\beta_c<\tfrac{1}{3}\big)
\]
is the worst-case probability that an $L$-sample exceeds the planar tolerance; the inequality is Hoeffding's bound, which also holds for sampling without replacement \cite{hoeffding1963probability}. Hence $\rho_L$ decays exponentially in the sample size $L$, and the number $\tilde M^-$ of agents that the analysis treats as bad is dominated by $M^-+\mathrm{Bin}(M^+,\rho_L)$. The agent-level term of Theorem~\ref{thm:decentral} stays below one whenever $\tilde M^-<D$, which holds when $\tilde M^-<\lceil|\mathbf{Z}|/3\rceil$. Table~\ref{tab:pl} gives exact values: $\rho_L$ is small for moderate $\beta_c$ but large near $\beta_c=1/3$, where larger samples are needed and safety otherwise rests on the data-adaptive tolerance $N_j^-<D_j^+$.

Combining Theorems~\ref{thm:safety-suffices} and~\ref{thm:decentral} gives the end-to-end guarantee of $\mathsf{DeFSM}$.
\begin{corollary}[Convergence of $\mathsf{DeFSM}$]\label{cor:e2e}
After $T=1/\eta$ rounds,
\[
F_{\Cg}(\mathbf{x}^{(T)})\ \ge\ (1-\bar\sigma)\big(1-1/\mathrm{e}\big)\,\mathrm{OPT}\;-\;2r\zeta\;-\;\frac{\eta\, m\, r^2}{2},
\]
where $\bar\sigma=\frac1T\sum_t\sigma_t$ and $\sigma_t$ is the minimum of 1 and the bound of Theorem~\ref{thm:decentral} in round $t$. Per round, $O(N+M)$ numbers are stored on-chain, independent of $d$, and each client downloads at most $D_t+2$ agent aggregates.
\end{corollary}
\begin{proof}
By Theorem~\ref{thm:decentral} every aggregate lies in $\mathcal{P}$ and its deficit is at most $\sigma_t$; Theorem~\ref{thm:safety-suffices} and the inequality $1-\mathrm{e}^{-(1-\bar\sigma)}\ge(1-\bar\sigma)(1-1/\mathrm{e})$ then apply. The costs follow from the protocol.
\end{proof}
Corollary~\ref{cor:e2e} guarantees graceful degradation under any attack: the guarantee depends on the adversary only through the deficit, and the deficit is bounded through counts and depths, never through distances that the adversary could manipulate. The quantity that $\mathsf{SafeGM}$ drives toward zero, and that we report empirically, is the realized deficit $\sigma_{\Cg}(\mathbf{w}^{(t)})$. It is computable in simulation by a linear program, because $\mathcal{P}$ is a polytope and ${\Cg}$ is known to the experimenter.

\section{Numerical Evaluation}\label{sec:eval}
We evaluate $\mathsf{DeFSM}$ numerically to validate its analysis and compare it with baselines. All simulation code, data, and results are available at \url{https://github.com/UMB-NCLab/defsm}. Here $\mathsf{DeFSM}$ denotes the architecture with its rule $\mathsf{SafeGM}$. We test four hypotheses.
\begin{itemize}
\item[H1] (Accuracy). Without any trusted party, $\mathsf{DeFSM}$ retains most of the quality of the geometric median at a trusted server, and stays close to the geometric median decentralized in an idealized setting.
\item[H2] (Efficiency). $\mathsf{DeFSM}$ needs much less communication than the straightforward decentralization where the agents run the robust aggregator in $\mathbb{R}^d$ and either agree on the result or let the clients reconstruct it.
\item[H3] (Robustness). $\mathsf{DeFSM}$ does not collapse, even under attacks designed against it, extreme data heterogeneity, and unknown Byzantine rates. In contrast, rate-dependent rules can collapse.
\item[H4] (Design parameters). The deficit does not depend on the number of clients and decreases with the sample size, as predicted, and the client download grows linearly with the number of agents.
\end{itemize}

\textbf{Baselines. } Agent-$X$ denotes aggregation rule $X$ (geometric median, trimmed mean, or mean) run by the agents of $\mathsf{DeFSM}$ in place of $\mathsf{SafeGM}$: it applies $X$ at both levels, with the same agents, samples, and bad agents, and deduplicates as $\mathsf{SafeGM}$ does. Agent-GM, with the geometric median, the most accurate robust rule in practice, is our main comparison; since the contract cannot compute it, we idealize it: the agents compute it and their vote always returns the good result. Agent-Trim, with the coordinate-wise trimmed mean, knows the true numbers of bad clients and agents unless stated otherwise. Agent-Mean is the non-robust reference. FSM-GM, the geometric median applied by a trusted central server to all client updates, is the centralized ideal reference. We also tested Krum, but its accuracy collapsed in every setting we tried, and so we omit it from the results.

\textbf{Metrics. } The \textit{quality} $Q=\frac{1}{N^+}\sum_{i\in {\Cg}}f_i(S^*)$ of the final solution $S^*$, obtained by pipage rounding (Section~\ref{sec:suffices}) and averaged over 10 roundings, is normalized so that 1 is the quality obtained by averaging the good clients' updates only and 0 that of a random subset. The safety \textit{deficit} $\sigma$ (Definition~\ref{def:safety}) of each round's aggregate is computed exactly by a linear program, since ${\Cg}$ is known in simulation. \textit{Communication} is the number of nonzero entries an agent and a client download per round.

\subsection{Evaluation Setting}\label{sec:setup}
A simulator runs the protocol round by round. We first run the \emph{default matrix}: all five attacks under all four combinations of Byzantine rates, with the other factors at their defaults. We then run ablations by varying one factor at a time: the heterogeneity, the numbers of clients and agents, the sample size, and the rule at the agent level. Table~\ref{tab:params} lists the defaults and the varied values.

\begin{wraptable}{r}{0.5\textwidth}
\vspace{-1.2em}\centering\footnotesize
\caption{Default values, and the values to which one factor at a time is varied.}
\label{tab:params}
\setlength{\tabcolsep}{2.5pt}
\begin{tabular}{lll}
\toprule
Factor & Default & Varied\\
\midrule
Datasets & CIFAR10, PathMNIST & ---\\
Ground set, budget & $d=200$, $r=10$ & ---\\
Clients & $N=60$ & $N\in\{200,500\}$\\
Heterogeneity & Dirichlet $\alpha=0.5$ & $\alpha=0.01$\\
Agents per round & $M=20$ & $M\in\{10,40\}$\\
Sample size & $L=20$ & $L\in\{10,40\}$\\
Byzantine rates & \multicolumn{2}{l}{all pairs of $\beta_c\in\{0.2,0.3\}$, $\beta_a\in\{0.1,0.3\}$}\\
Client attacks & \multicolumn{2}{l}{all five (see Attacks)}\\
Rate told Agent-Trim & true & true $\pm0.1$\\
Agent-level rule & $\mathsf{SafeGM}$ point & mean\\
Step size, rounds & $\eta=0.01$, $\ge100$ rounds & ---\\
Seeds & 3 & 10 ($\alpha$, $N$), 5 ($M$, $L$)\\
\bottomrule
\end{tabular}
\vspace{-1em}
\end{wraptable}
\textbf{Task and data. } Clients jointly select $r=10$ images that best represent their image collections. We use two real-world datasets of 60K images each: CIFAR10 \cite{krizhevsky2009learning} (10 object classes) and PathMNIST \cite{DBLP:journals/corr/abs-2110-14795} (9 classes of pathology tissue), both represented by standardized 512-dimensional ResNet18 features \cite{he2016deep} pretrained on ImageNet. In each run, the ground set $E$ consists of $d=200$ images drawn uniformly at random. Each client holds 40 images drawn from a client-specific class mixture sampled from a Dirichlet distribution with concentration $\alpha=0.5$, so the data are non-iid. Client $i$'s valuation is the facility-location function $f_i(S)=\sum_{p\in \mathcal{X}_i}\max_{s\in S}\exp\!\big(-\|p-s\|^2/(0.5h)^2\big)$, where $h$ is the median distance between ground-set images; it is monotone submodular. Clients estimate the gradient of their multilinear extension from 8 random samples and return the binary top-$r$ vector. Each run performs at least $1/\eta=100$ rounds, the horizon of Theorem~\ref{thm:safety-suffices}, and stops once the global top-$r$ set is unchanged for 15 rounds (at most 300); most stop at 100. Beyond $1/\eta$ rounds, $\mathbf{x}$ may leave $\mathcal{P}$; it is then clipped to $[0,1]^d$ and rescaled to sum $r$ before rounding.

\textbf{Attacks. } All bad client updates are valid binary vectors with $r$ ones, and so the validity filter cannot reject them. We use five client attacks:
\begin{enumerate}
\item \emph{Include}: every bad client submits the $r$ elements ranked lowest by the current global solution, the strongest coordinated push toward a poor solution;
\item \emph{Camouflage}: half of a good client's choice plus the lowest-ranked elements, identical across bad clients, to look partly legitimate while still promoting poor elements;
\item \emph{Near-Include}: Include with one element swapped per bad client, so that bad updates are distinct and deduplication cannot remove them; it targets $\mathsf{SafeGM}$ specifically and is the worst case for its deduplication;
\item \emph{Mimic}: each bad client copies a different good client's update and replaces half of its elements with lowest-ranked ones, to lie close to good updates in every projection, where depth cannot separate them; it also targets $\mathsf{SafeGM}$ and is the worst case for its depth test;
\item \emph{Random}: each bad client submits $r$ random elements, an uncoordinated adversary that only adds noise.
\end{enumerate}
A bad agent, in every method, posts as its aggregate the Include vector with one element swapped, distinct for each bad agent so that deduplication cannot merge them. This is the strongest push toward poor elements; by Theorem~\ref{thm:decentral}, other behaviors cannot gain more than their count allows.

\subsection{Numerical Results}

\subsubsection{Accuracy (H1)}
\begin{figure}[tbp]
\begin{minipage}[t]{0.44\textwidth}\vspace{0pt}\centering
\includegraphics[width=\linewidth]{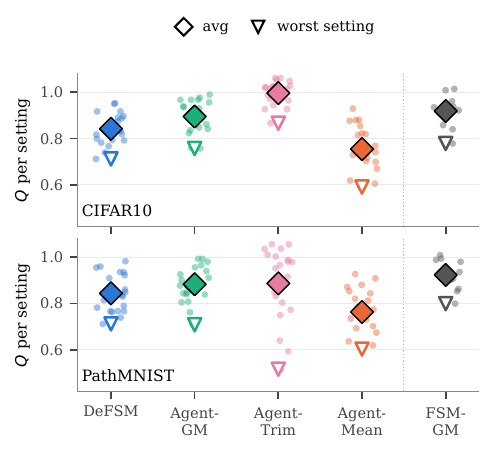}
\caption{Accuracy summary. Small points: one per setting of the default matrix (mean over 3 seeds); 10 settings for FSM-GM. Diamond: average; open triangle: worst setting.}
\label{fig:h1}
\end{minipage}\hfill
\begin{minipage}[t]{0.53\textwidth}\vspace{0pt}\centering\footnotesize
\captionof{table}{Reliability over all 20 attack settings (5 attacks $\times$ $\beta_c\in\{0.2,0.3\}$ $\times$ $\beta_a\in\{0.1,0.3\}$), 3 seeds, $\alpha=0.5$. Agent-Trim is told the true Byzantine rates unless stated; FSM-GM has 10 settings ($\beta_c$ only). Worst $\sigma$: largest deficit of any run. Worst run / worst setting / avg: smallest $Q$ of any run, smallest average $Q$ of any setting, average $Q$. Bold: best in column.}
\label{tab:summary}
\setlength{\tabcolsep}{2.0pt}
\begin{tabular}{lcccccccc}
\toprule
& \multicolumn{4}{c}{CIFAR10} & \multicolumn{4}{c}{PathMNIST}\\
\cmidrule(lr){2-5}\cmidrule(lr){6-9}
Method & \makecell{worst\\$\sigma$} & \makecell{worst\\run} & \makecell{worst\\setting} & \makecell{avg\\$Q$} & \makecell{worst\\$\sigma$} & \makecell{worst\\run} & \makecell{worst\\setting} & \makecell{avg\\$Q$}\\
\midrule
$\mathsf{DeFSM}$ & 0.41 & 0.63 & 0.71 & 0.84 & 0.44 & 0.61 & \textbf{0.71} & 0.84\\
\midrule
Agent-GM & \textbf{0.35} & 0.64 & 0.76 & 0.89 & \textbf{0.39} & \textbf{0.66} & 0.71 & 0.88\\
Agent-Trim & 1.00 & \textbf{0.78} & \textbf{0.87} & \textbf{1.00} & 1.00 & 0.36 & 0.52 & \textbf{0.89}\\
Agent-Mean & 0.50 & 0.39 & 0.59 & 0.75 & 0.50 & 0.47 & 0.60 & 0.76\\
\midrule\multicolumn{9}{l}{\emph{Centralized ideal reference}}\\
FSM-GM & 0.27 & 0.73 & 0.78 & 0.92 & 0.30 & 0.73 & 0.80 & 0.92\\
\midrule\multicolumn{9}{l}{\emph{Agent-Trim told a wrong Byzantine rate (true $\pm0.1$)}}\\
Agent-Trim, $-0.1$ & 1.00 & 0.68 & 0.79 & 0.98 & 0.99 & 0.47 & 0.57 & 0.87\\
Agent-Trim, $+0.1$ & 1.00 & 0.60 & 0.75 & 0.95 & 1.00 & 0.06 & 0.35 & 0.78\\
\bottomrule
\end{tabular}
\end{minipage}
\end{figure}
\begin{figure}[tbp]
\centering
\includegraphics[width=\textwidth]{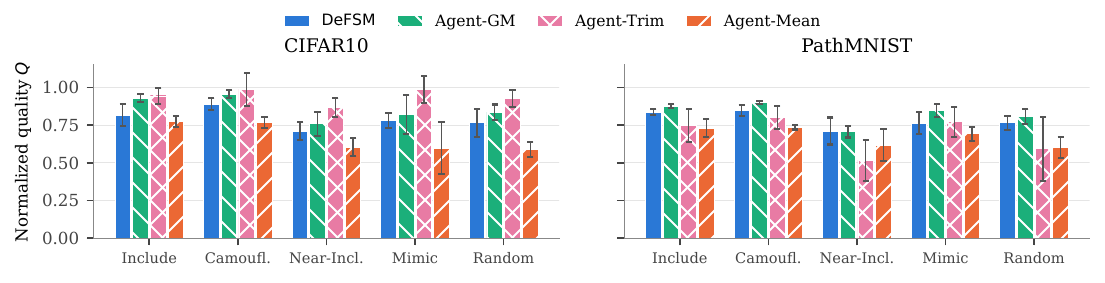}
\caption{Normalized quality $Q$ in the hardest setting ($\beta_c=\beta_a=0.3$), mean and standard deviation over 3 seeds.}
\label{fig:quality}
\end{figure}
Fig.~\ref{fig:h1} summarizes accuracy over the default matrix, and Table~\ref{tab:summary} gives the numbers. $\mathsf{DeFSM}$ and Agent-GM sit close together, with similar worst settings; Agent-Trim is higher on average but spreads widely on PathMNIST, and Agent-Mean is lowest. FSM-GM reaches 0.92 on both datasets and $\mathsf{DeFSM}$ 0.84, about 91\% of it without any trusted party. Agent-GM, decentralized in an idealized setting, reaches 0.88--0.89, about 96\%, with a worst run similar to that of $\mathsf{DeFSM}$ (0.64--0.66 against 0.61--0.63).

\textbf{Hardest setting. } Fig.~\ref{fig:quality} shows each attack at $\beta_c=\beta_a=0.3$. On PathMNIST, $\mathsf{DeFSM}$ ties with Agent-GM under Near-Include and is within 0.04--0.09 of it under the other attacks, ahead of Agent-Trim except under Mimic. On CIFAR10, it ranks third under every attack, behind Agent-Trim and Agent-GM, by 0.04--0.11 from Agent-GM. Agent-Mean is clearly behind in every case.

\begin{figure}[tbp]
\centering
\includegraphics[width=\textwidth]{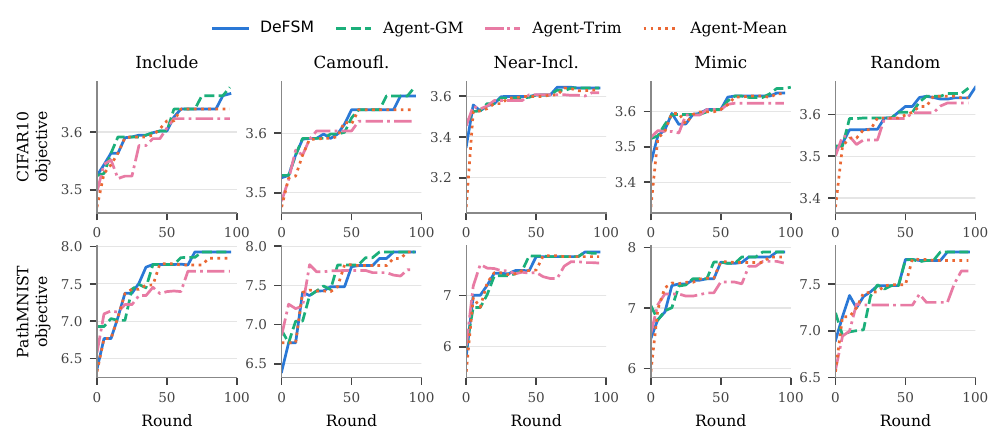}
\caption{Good clients' objective of the top-$r$ set of $\mathbf{x}$ per round ($\beta_c=\beta_a=0.3$, mean over 3 seeds).}
\label{fig:convergence}
\end{figure}
\textbf{Convergence. } Under every attack and on both datasets, $\mathsf{DeFSM}$ converges as fast as Agent-GM and Agent-Mean to a similar objective (Fig.~\ref{fig:convergence}). These results support H1: $\mathsf{DeFSM}$ gives up a few points of quality, about 0.05 against Agent-GM, in exchange for removing the trusted party.

\subsubsection{Efficiency (H2)}
\begin{figure}[tbp]
\begin{minipage}[t]{0.48\textwidth}\vspace{0pt}\centering
\includegraphics[width=\linewidth]{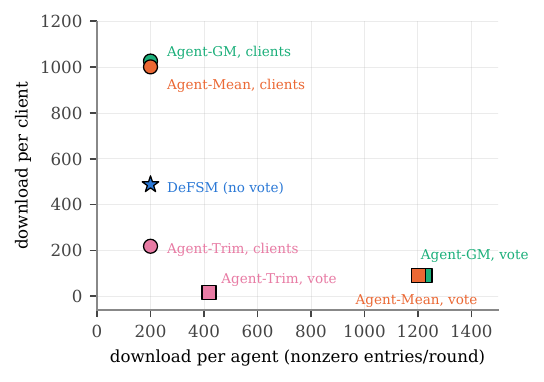}
\caption{Efficiency summary: download per agent and per client in each round, averaged over all settings. Baselines are shown with the clients or the agents (vote) computing the global aggregate.}
\label{fig:h2}
\end{minipage}\hfill
\begin{minipage}[t]{0.48\textwidth}\vspace{0pt}\centering
\includegraphics[width=\linewidth]{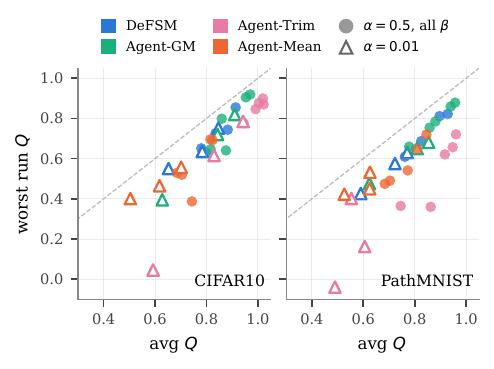}
\caption{Stability summary: average against worst-run quality for each method and attack, over all Byzantine rates at $\alpha=0.5$ (filled, 3 seeds per setting) and at extreme heterogeneity $\alpha=0.01$, $\beta_c=\beta_a=0.3$ (open, 10 seeds). Dashed: worst run equal to average.}
\label{fig:h3}
\end{minipage}
\end{figure}
Fig.~\ref{fig:h2} places each method on the plane of agent and client download. In $\mathsf{DeFSM}$, the contract decides from $2M=40$ numbers, an agent downloads only its $L$ client updates (200 entries), and a client downloads 9.3 agent aggregates on average (487 entries), within the bound $D+2$ of Corollary~\ref{cor:e2e}. Since the contract cannot compute Agent-GM, either each agent also downloads all agent aggregates (1{,}226 entries, six times more) and the agents vote, or each client downloads all 19.9 of them (1{,}026 entries, twice as many). Agent-Mean costs about the same. Agent-Trim is cheaper because its agent aggregates are nearly binary, but it is the least reliable robust option, as we show next. In every method, on-chain storage is independent of $d$. These results support H2.

\subsubsection{Robustness (H3)}
\begin{figure}[tbp]
\begin{minipage}[t]{0.46\textwidth}\vspace{0pt}\centering
\includegraphics[width=\linewidth]{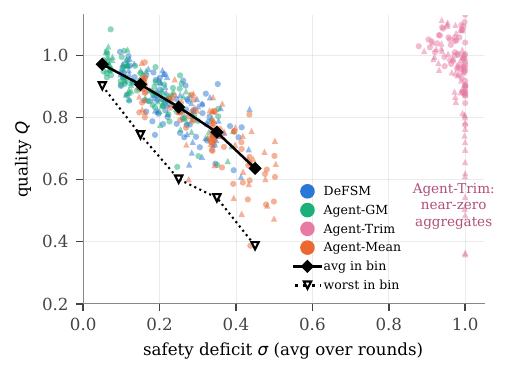}
\caption{Safety deficit against quality for every run of the default matrix (3 seeds; circles CIFAR10, triangles PathMNIST). Lines: average and worst quality of the $\mathsf{DeFSM}$, Agent-GM, and Agent-Mean runs per deficit bin of width 0.1.}
\label{fig:deficit}
\end{minipage}\hfill
\begin{minipage}[t]{0.51\textwidth}\vspace{0pt}\centering\footnotesize
\captionof{table}{Extreme heterogeneity (Dirichlet $\alpha=0.01$, $\beta_c=\beta_a=0.3$), 10 seeds: avg / worst normalized quality $Q$. Bold: best worst case.}
\label{tab:hetero}
\setlength{\tabcolsep}{1.6pt}
\begin{tabular}{llcccc}
\toprule
Dataset & Attack & $\mathsf{DeFSM}$ & Agent-GM & Agent-Trim & Agent-Mean\\
\midrule
CIFAR10 & Include & 0.85 / 0.75 & 0.91 / \textbf{0.82} & 0.83 / 0.62 & 0.70 / 0.56\\
 & Near-I. & 0.65 / \textbf{0.55} & 0.63 / 0.39 & 0.59 / 0.05 & 0.50 / 0.40\\
 & Mimic & 0.78 / 0.64 & 0.84 / 0.72 & 0.94 / \textbf{0.79} & 0.62 / 0.47\\
\midrule
PathMNIST & Include & 0.77 / 0.63 & 0.85 / \textbf{0.68} & 0.55 / 0.40 & 0.62 / 0.45\\
 & Near-I. & 0.59 / 0.43 & 0.62 / \textbf{0.48} & 0.49 / $-0.04$ & 0.53 / 0.42\\
 & Mimic & 0.72 / 0.58 & 0.81 / \textbf{0.65} & 0.61 / 0.16 & 0.63 / 0.53\\
\bottomrule
\end{tabular}
\end{minipage}
\end{figure}
\begin{figure}[tbp]
\centering
\includegraphics[width=\textwidth]{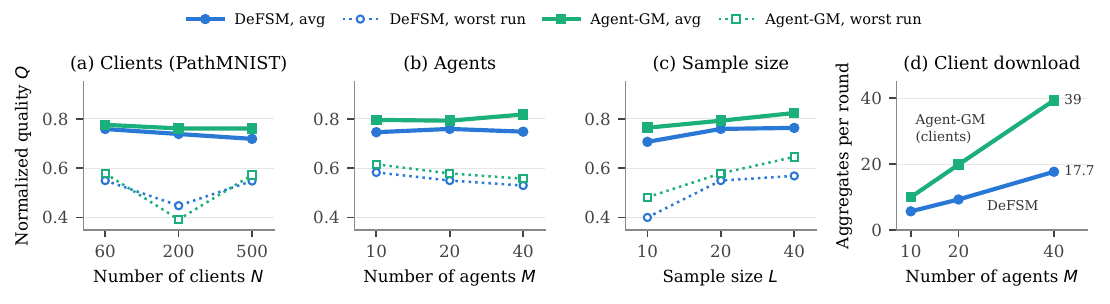}
\caption{Effect of (a) the number of clients $N$ (PathMNIST, 10 seeds), (b) the number of agents $M$, and (c) the sample size $L$ (both datasets, 5 seeds) on the average and worst normalized quality of $\mathsf{DeFSM}$ and Agent-GM, at $\beta_c=\beta_a=0.3$ under Near-Include and Mimic, varying one parameter at a time. (d) Agent aggregates a client downloads per round.}
\label{fig:ablation}
\end{figure}
\textbf{Stability at a glance. } In Fig.~\ref{fig:h3}, the points of $\mathsf{DeFSM}$ and Agent-GM stay together in the upper band at both heterogeneity levels. Agent-Mean sits lower, and Agent-Trim reaches high averages but has the lowest worst runs, down to $-0.04$ at $\alpha=0.01$.

\textbf{Attacks and Byzantine rates. } Over the default matrix (Table~\ref{tab:summary}), the worst run of $\mathsf{DeFSM}$ is 0.63 on CIFAR10 and 0.61 on PathMNIST. The other rules each fail somewhere. Agent-Mean loses up to 0.6 of the good clients' quality. Agent-Trim's worst run on PathMNIST is 0.36, although it knows the exact Byzantine rates. When the rate it is told is off by $0.1$, its worst run on PathMNIST falls to 0.06. $\mathsf{DeFSM}$ and Agent-GM need no such knowledge.

\textbf{Deficit drives quality. } Theorem~\ref{thm:safety-suffices} predicts that the quality degrades only through the safety deficit. Fig.~\ref{fig:deficit} confirms it across all runs of $\mathsf{DeFSM}$, Agent-GM, and Agent-Mean: the lower the deficit, the higher the quality, with a rank correlation of $-0.86$ ($-0.75$ to $-0.85$ within each method). The average and worst quality fall steadily, from 0.97 and 0.90 below $\sigma=0.1$ to 0.64 and 0.39 at $\sigma=0.4$--$0.5$. The condition is only sufficient: Agent-Trim's deficit is near 1 because its aggregates are nearly zero, yet its quality varies widely. The deficit of $\mathsf{DeFSM}$ never exceeds 0.44.

\textbf{Extreme heterogeneity. } Table~\ref{tab:hetero} sets the Dirichlet concentration to 0.01, so that most clients hold a single class and good updates differ widely; we use 10 seeds here. Agent-GM or Agent-Trim is the most accurate on average except under Near-Include on CIFAR10, where $\mathsf{DeFSM}$ leads in both average (0.65 against 0.63) and worst run (0.55 against 0.39). Over all our experiments with the default sample size $L=20$ (Tables~\ref{tab:summary} and~\ref{tab:hetero}, and Fig.~\ref{fig:ablation}(a)), the worst run of $\mathsf{DeFSM}$ is 0.43, against 0.39 for Agent-GM and Agent-Mean and $-0.04$ for Agent-Trim; $\mathsf{DeFSM}$ and Agent-GM each have 4 runs out of 240 below 0.5, Agent-Mean 22, and Agent-Trim 27.

\textbf{Deep point at the agent level. } Replacing the $\mathsf{SafeGM}$ point at the agent level by the mean of the distinct agent aggregates, with $\mathsf{SafeGM}$ unchanged inside each agent, lowers the quality at $\beta_c=\beta_a=0.3$ from 0.79 to 0.67 on CIFAR10 and from 0.79 to 0.68 on PathMNIST, and raises the deficit from 0.29--0.31 to 0.41--0.42. The mean has the smaller worst-case bound (Section~\ref{sec:decguarantee}), but the deep point gives bad agent aggregates a much smaller share in practice. These results support H3: $\mathsf{DeFSM}$ degrades gracefully, as stably as Agent-GM, while Agent-Mean and Agent-Trim collapse in some settings.

\subsubsection{Design Parameters (H4)}
\begin{figure}[tbp]
\centering
\includegraphics[width=\textwidth]{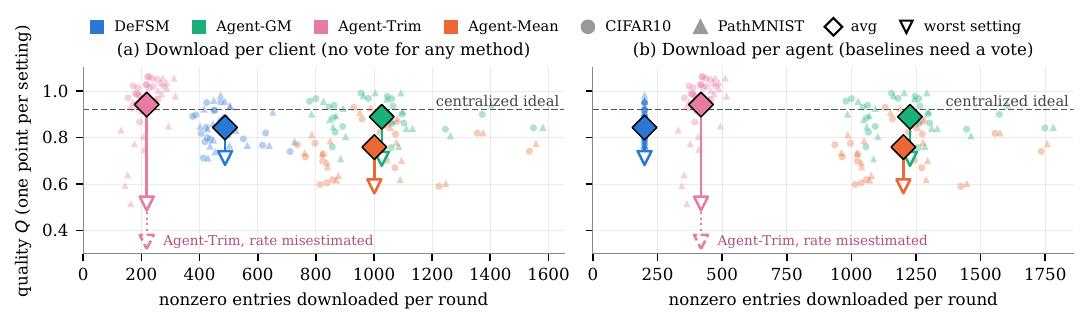}
\caption{Quality against download per round. Small points: one per setting of the default matrix (mean over 3 seeds). Diamond: average; open triangle: worst setting. Baselines compute the global aggregate at (a) the clients or (b) the agents, with a vote. Dotted: Agent-Trim told a Byzantine rate 0.1 too high. Dashed: FSM-GM.}
\label{fig:tradeoff}
\end{figure}
\textbf{Number of clients. } Fig.~\ref{fig:ablation}(a) increases the number of clients to 200 and 500 with the sample size fixed at $L=20$. The deficit of $\mathsf{DeFSM}$ stays essentially unchanged (0.42--0.43 under Near-Include and 0.31--0.34 under Mimic), as the analysis predicts: its tolerance depends on the Byzantine client rate and $L$, not on $N$. Its quality decreases slightly with $N$, since each agent sees a smaller fraction of the clients, and stays within 0.05 of Agent-GM on average; its worst run is higher than Agent-GM's at $N=200$ (0.45 against 0.39).

\textbf{Number of agents. } The quality of $\mathsf{DeFSM}$ hardly changes with $M$ (Fig.~\ref{fig:ablation}(b)), while the number of agent aggregates a client downloads grows linearly with $M$ (5.7, 9.3, and 17.7; Fig.~\ref{fig:ablation}(d)), since the coefficients spread over at least $D\ge\lceil M/3\rceil$ of them, whereas Agent-GM at the clients needs all of them (10, 20, and 39). $M$ therefore only needs to be large enough to keep the agent-level term of Theorem~\ref{thm:decentral} small.

\textbf{Sample size. } A larger sample lowers the deficit of $\mathsf{DeFSM}$ (0.42, 0.37, and 0.34 for $L=10$, 20, and 40), as the analysis of $\rho_L$ predicts (Table~\ref{tab:pl}). Small samples hurt $\mathsf{DeFSM}$ more than Agent-GM (Fig.~\ref{fig:ablation}(c)): with $L=10$, its worst run drops to 0.40, against 0.48 for Agent-GM, because an agent's sample then exceeds the planar tolerance more often. These results support H4 and suggest $L\ge20$.

\subsubsection{Overall Picture}

Fig.~\ref{fig:tradeoff} combines accuracy and cost, with one point per method and setting of the default matrix. $\mathsf{DeFSM}$ needs about half the client download of Agent-GM and Agent-Mean, and a sixth of their agent download when they vote, and its points stay in a narrow band: its worst setting matches Agent-GM's and is well above Agent-Mean's. Agent-Trim is cheaper and more accurate on average, but spreads widest, and a misestimated rate drops its worst setting to 0.35. Only $\mathsf{DeFSM}$ combines low cost, stable quality, and no need to know the Byzantine rates. As for the design parameters, $\mathsf{DeFSM}$ scales to many clients, the number of agents trades only client download, and the sample size is the lever on robustness.

\section{Conclusions}\label{sec:conclusion}
Submodular maximization is a core tool in machine learning, and its federated version must work where a trusted server is infeasible. Removing the server creates a tension between decentralization, efficiency, and security. Our $\mathsf{DeFSM}$ architecture resolves it with a lightweight smart contract and a pool of agents that need no consensus, and its aggregation rule $\mathsf{SafeGM}$ is provably robust. We learned that robustness need not be judged by distances in the full space: a depth-certified decision in two dimensions bounds the influence of bad updates by their count, this bound controls the solution quality, and in practice it comes close to the geometric median at a fraction of the communication, without consensus or knowledge of the Byzantine rates.
The impact reaches beyond FSM. It makes robust federated optimization deployable on public blockchains, where anyone can serve as an agent and no party must be trusted; hospitals, for example, could jointly curate the most informative medical images without a common coordinator. It suggests a general pattern for verifiable decentralized computing: off-chain workers do the high-dimensional work, and the chain decides on a low-dimensional certificate. 
It also brings combinatorial tools such as Tukey depth, long confined to very low dimension, to high-dimensional systems, with guarantees that hold by counting rather than by distance.
In future work, we will implement $\mathsf{DeFSM}$ on a real blockchain and measure its gas fees, storage and communication overheads, and latency per round to assess its feasibility in practice.

\section*{CRediT authorship contribution statement}
\textbf{Duc A. Tran:} Conceptualization, Methodology, Writing -- original draft, Writing -- review \& editing. \textbf{Dung Truong:} Conceptualization, Software, Data curation. \textbf{Duy Le:} Conceptualization, Software.

\section*{Declaration of competing interest}
The authors declare that they have no known competing financial interests or personal relationships that could have appeared to influence the work reported in this paper.

\section*{Funding}
This research did not receive any specific grant from funding agencies in the public, commercial, or not-for-profit sectors.

\section*{Data availability}
All simulation code, data, and results are available at \url{https://github.com/UMB-NCLab/defsm}.

\section*{Declaration of generative AI and AI-assisted technologies in the writing process}
During the preparation of this work we used AI as a plotting tool for the illustrative diagram in Figure 1 and the evaluation figures, where we input our own conceptual description and numerical values. We also used Claude AI in part of our coding for the evaluation study, and to improve the language and presentation of the manuscript. After all this AI help, we reviewed and edited the content to ensure correctness, and take full responsibility for the content of this article. We confirm that all the main, intellectual contents (ideas, literature review, algorithms, analysis, evaluation) are our own.

\bibliographystyle{elsarticle-num}
\bibliography{bibliography,misc,thispaper_defed,tran,thispaper_peerfed,thisproposal,revision}

@article{DBLP:journals/corr/abs-2110-14795,
  author       = {Jiancheng Yang and Rui Shi and Donglai Wei and Zequan Liu and Lin Zhao and Bilian Ke and Hanspeter Pfister and Bingbing Ni},
  title        = {{MedMNIST} v2 -- A Large-Scale Lightweight Benchmark for {2D} and {3D} Biomedical Image Classification},
  journal      = {Scientific Data},
  volume       = {10},
  pages        = {41},
  year         = {2023},
  doi          = {10.1038/s41597-022-01721-8}
}

@INPROCEEDINGS{10074277,
  author={Mallah, Ranwa Al and López, David and Halabi, Talal},
  booktitle={2023 International Conference on Computing, Networking and Communications (ICNC)}, 
  title={Blockchain-enabled Efficient and Secure Federated Learning in {IoT} and Edge Computing Networks}, 
  year={2023},
  pages={511-515},
  doi={10.1109/ICNC57223.2023.10074277}}

@article{10.1137/080733991,
author = {Calinescu, Gruia and Chekuri, Chandra and P\'{a}l, Martin and Vondr\'{a}k, Jan},
title = {Maximizing a Monotone Submodular Function Subject to a Matroid Constraint},
year = {2011},
issue_date = {November 2011},
publisher = {Society for Industrial and Applied Mathematics},
address = {USA},
volume = {40},
number = {6},
issn = {0097-5397},
doi = {10.1137/080733991},
journal = {SIAM J. Comput.},
month = dec,
pages = {1740–1766},
numpages = {27}
}

@inproceedings{10.1145/1374376.1374389,
author = {Vondrak, Jan},
title = {Optimal approximation for the submodular welfare problem in the value oracle model},
year = {2008},
isbn = {9781605580470},
publisher = {Association for Computing Machinery},
address = {New York, NY, USA},
doi = {10.1145/1374376.1374389},
booktitle = {Proceedings of the Fortieth Annual ACM Symposium on Theory of Computing},
pages = {67–74},
numpages = {8},
location = {Victoria, British Columbia, Canada},
series = {STOC '08}
}

@misc{vendrell2024optimalitygapdecentralizedsubmodular,
      title={Optimality Gap of Decentralized Submodular Maximization under Probabilistic Communication}, 
      author={Joan Vendrell and Solmaz Kia},
      year={2024},
      eprint={2409.09979},
      archivePrefix={arXiv},
      primaryClass={cs.MA},
      url={https://arxiv.org/abs/2409.09979}, 
}

@article{10.5555/2946645.3053517,
author = {Mirzasoleiman, Baharan and Karbasi, Amin and Sarkar, Rik and Krause, Andreas},
title = {Distributed submodular maximization},
year = {2016},
issue_date = {January 2016},
publisher = {JMLR.org},
volume = {17},
number = {1},
issn = {1532-4435},
journal = {J. Mach. Learn. Res.},
month = jan,
pages = {8330–8373},
numpages = {44}
}

@InProceedings{pmlr-v37-barbosa15,
  title = 	 {The Power of Randomization: Distributed Submodular Maximization on Massive Datasets},
  author = 	 {Barbosa, Rafael and Ene, Alina and Nguyen, Huy and Ward, Justin},
  booktitle = 	 {Proceedings of the 32nd International Conference on Machine Learning},
  pages = 	 {1236--1244},
  year = 	 {2015},
  editor = 	 {Bach, Francis and Blei, David},
  volume = 	 {37},
  series = 	 {Proceedings of Machine Learning Research},
  address = 	 {Lille, France},
  month = 	 {07--09 Jul},
  publisher =    {PMLR}
}

@article{10.1007/BF01588971,
author = {Nemhauser, G. L. and Wolsey, L. A. and Fisher, M. L.},
title = {An analysis of approximations for maximizing submodular set functions--{I}},
year = {1978},
issue_date = {December  1978},
publisher = {Springer-Verlag},
address = {Berlin, Heidelberg},
volume = {14},
number = {1},
issn = {0025-5610},
doi = {10.1007/BF01588971},
journal = {Math. Program.},
month = dec,
pages = {265–294},
numpages = {30}
}

@inproceedings{DBLP:conf/icml/Rafiey24,
  author       = {Akbar Rafiey},
  title        = {Decomposable Submodular Maximization in Federated Setting},
  booktitle    = {Forty-first International Conference on Machine Learning, {ICML} 2024,
                  Vienna, Austria, July 21-27, 2024},
  publisher    = {OpenReview.net},
  year         = {2024},
  bibsource    = {dblp computer science bibliography, https://dblp.org}
}

@ARTICLE{10286024,
  author={Wang, Yanhao and Zhou, Tianchen and Chen, Cen and Wang, Yinggui},
  journal={IEEE Internet of Things Journal}, 
  title={Federated Submodular Maximization With Differential Privacy}, 
  year={2024},
  volume={11},
  number={2},
  pages={1827-1839},
  doi={10.1109/JIOT.2023.3324801}}

@ARTICLE{9293091,
  author={Li, Yuzheng and Chen, Chuan and Liu, Nan and Huang, Huawei and Zheng, Zibin and Yan, Qiang},
  journal={IEEE Network}, 
  title={A Blockchain-Based Decentralized Federated Learning Framework with Committee Consensus}, 
  year={2021},
  volume={35},
  number={1},
  pages={234-241},
  doi={10.1109/MNET.011.2000263}}

@inproceedings{10.1145/3579654.3579700,
author = {Lin, Zhihao and Wu, Zhijie and Zhu, Jia and Huang, Jin},
title = {A Blockchain-Based Decentralized Federated Learning Framework with Dual-Committees Consensus},
booktitle={ACAI '22: Proceedings of the 2022 5th International Conference on Algorithms, Computing and Artificial Intelligence},
year = {2023},
isbn = {9781450398336},
publisher = {Association for Computing Machinery},
address = {New York, NY, USA},
doi = {10.1145/3579654.3579700},
articleno = {43},
numpages = {6},
location = {Sanya, China},
series = {ACAI '22}
}

@INPROCEEDINGS{10001370,
  author={Yang, Zheng and Li, Gaolei and Wu, Jun and Yang, Wu},
  booktitle={GLOBECOM 2022 - 2022 IEEE Global Communications Conference}, 
  title={Propagable Backdoors over Blockchain-based Federated Learning via Sample-Specific Eclipse}, 
  year={2022},
  volume={},
  number={},
  pages={2579-2584},
  doi={10.1109/GLOBECOM48099.2022.10001370}}

@article{nguyen2023backdoor,
  title={Backdoor Attacks and Defenses in Federated Learning: Survey, Challenges and Future Research Directions}, 
      author={Thuy Dung Nguyen and Tuan Nguyen and Phi Le Nguyen and Hieu H. Pham and Khoa Doan and Kok-Seng Wong},
      year={2024},
      journal={Engineering Applications of Artificial Intelligence}
}

@article{pillutla2022robust,
  title={Robust aggregation for federated learning},
  author={Pillutla, Krishna and Kakade, Sham M and Harchaoui, Zaid},
  journal={IEEE Transactions on Signal Processing},
  volume={70},
  pages={1142--1154},
  year={2022},
  publisher={IEEE}
}

@inproceedings{Yin2018ByzantineRobustDL,
  title={Byzantine-Robust Distributed Learning: Towards Optimal Statistical Rates},
  author={Dong Yin and Yudong Chen and Kannan Ramchandran and Peter L. Bartlett},
  booktitle={Proceedings of the 35th International Conference on Machine Learning (ICML)},
  series={Proceedings of Machine Learning Research},
  volume={80},
  pages={5650--5659},
  year={2018}
}

@inproceedings{Madill2022ScaleSFLAS,
  title={{ScaleSFL}: A Sharding Solution for Blockchain-Based Federated Learning},
  author={Evan W. R. Madill and Ben Nguyen and Carson Kai-Sang Leung and Sara Rouhani},
  booktitle={Proceedings of the Fourth ACM International Symposium on Blockchain and Secure Critical Infrastructure (BSCI)},
  pages={95--106},
  doi={10.1145/3494106.3528680},
  year={2022}
}

@techreport{krizhevsky2009learning,
  author = {Krizhevsky, Alex and Hinton, Geoffrey},
  title = {Learning Multiple Layers of Features from Tiny Images},
  institution = {University of Toronto, Technical Report CRG-TR-09-1},
  year = {2009},
  url = {https://www.cs.toronto.edu/~kriz/learning-features-2009-TR.pdf}
}

@inproceedings{DBLP:conf/mlsys/LiSZSTS20,
  author    = {Tian Li and
               Anit Kumar Sahu and
               Manzil Zaheer and
               Maziar Sanjabi and
               Ameet Talwalkar and
               Virginia Smith},
  editor    = {Inderjit S. Dhillon and
               Dimitris S. Papailiopoulos and
               Vivienne Sze},
  title     = {Federated Optimization in Heterogeneous Networks},
  booktitle = {Proceedings of Machine Learning and Systems 2020, MLSys 2020, Austin,
               TX, USA, March 2-4, 2020},
  publisher = {mlsys.org},
  year      = {2020},
  bibsource = {dblp computer science bibliography, https://dblp.org}
}

@inproceedings{10.5555/296806.296824,
author = {Castro, Miguel and Liskov, Barbara},
title = {Practical Byzantine Fault Tolerance},
year = {1999},
isbn = {1880446391},
publisher = {USENIX Association},
address = {USA},
booktitle = {Proceedings of the Third Symposium on Operating Systems Design and Implementation},
pages = {173–186},
numpages = {14},
location = {New Orleans, Louisiana, USA},
series = {OSDI '99}
}

@article{10.1145/357172.357176,
author = {Lamport, Leslie and Shostak, Robert and Pease, Marshall},
title = {The Byzantine Generals Problem},
year = {1982},
issue_date = {July 1982},
publisher = {Association for Computing Machinery},
address = {New York, NY, USA},
volume = {4},
number = {3},
issn = {0164-0925},
url = {https://doi.org/10.1145/357172.357176},
doi = {10.1145/357172.357176},
journal = {ACM Trans. Program. Lang. Syst.},
month = {jul},
pages = {382–401},
numpages = {20}
}

@inproceedings{DBLP:conf/iclr/LiHYWZ20,
  author    = {Xiang Li and
               Kaixuan Huang and
               Wenhao Yang and
               Shusen Wang and
               Zhihua Zhang},
  title     = {On the Convergence of FedAvg on Non-IID Data},
  booktitle = {8th International Conference on Learning Representations, {ICLR} 2020,
               Addis Ababa, Ethiopia, April 26-30, 2020},
  year      = {2020},
  bibsource = {dblp computer science bibliography, https://dblp.org}
}

@article{megiddo1984linear,
  author  = {Megiddo, Nimrod},
  title   = {Linear Programming in Linear Time When the Dimension Is Fixed},
  journal = {Journal of the ACM},
  volume  = {31},
  number  = {1},
  pages   = {114--127},
  year    = {1984}
}

@article{rado1947theorem,
  author  = {Rado, Richard},
  title   = {A Theorem on General Measure},
  journal = {Journal of the London Mathematical Society},
  volume  = {s1-21},
  number  = {4},
  pages   = {291--300},
  year    = {1946}
}

@article{hoeffding1963probability,
  author  = {Hoeffding, Wassily},
  title   = {Probability Inequalities for Sums of Bounded Random Variables},
  journal = {Journal of the American Statistical Association},
  volume  = {58},
  number  = {301},
  pages   = {13--30},
  year    = {1963}
}

@inproceedings{he2016deep,
  author    = {He, Kaiming and Zhang, Xiangyu and Ren, Shaoqing and Sun, Jian},
  title     = {Deep Residual Learning for Image Recognition},
  booktitle = {Proceedings of the IEEE Conference on Computer Vision and Pattern Recognition (CVPR)},
  pages     = {770--778},
  year      = {2016}
}

@article{feige1998threshold,
  author  = {Feige, Uriel},
  title   = {A Threshold of $\ln n$ for Approximating Set Cover},
  journal = {Journal of the ACM},
  volume  = {45},
  number  = {4},
  pages   = {634--652},
  year    = {1998}
}

@inproceedings{chan2004optimal,
  title={An optimal randomized algorithm for maximum {Tukey} depth},
  author={Chan, Timothy M},
  booktitle={Proceedings of the 15th Annual ACM-SIAM Symposium on Discrete Algorithms (SODA)},
  pages={430--436},
  year={2004}
}

@inproceedings{10.1145/2484239.2484256,
author = {Vaidya, Nitin H. and Garg, Vijay K.},
title = {{B}yzantine vector consensus in complete graphs},
year = {2013},
isbn = {9781450320658},
publisher = {Association for Computing Machinery},
address = {New York, NY, USA},
doi = {10.1145/2484239.2484256},
booktitle = {Proceedings of the 2013 ACM Symposium on Principles of Distributed Computing},
pages = {65–73},
numpages = {9},
location = {Montr\'{e}al, Qu\'{e}bec, Canada},
series = {PODC '13}
}

@inproceedings{10.1145/2488608.2488657,
author = {Mendes, Hammurabi and Herlihy, Maurice},
title = {Multidimensional approximate agreement in {B}yzantine asynchronous systems},
year = {2013},
isbn = {9781450320290},
publisher = {Association for Computing Machinery},
address = {New York, NY, USA},
doi = {10.1145/2488608.2488657},
booktitle = {Proceedings of the Forty-Fifth Annual ACM Symposium on Theory of Computing},
pages = {391–400},
numpages = {10},
location = {Palo Alto, California, USA},
series = {STOC '13}
}

@inproceedings{DBLP:conf/icml/MhamdiGR18,
  author       = {El Mahdi {El Mhamdi} and
                  Rachid Guerraoui and
                  S{\'{e}}bastien Rouault},
  editor       = {Jennifer G. Dy and
                  Andreas Krause},
  title        = {The Hidden Vulnerability of Distributed Learning in {B}yzantium},
  booktitle    = {Proceedings of the 35th International Conference on Machine Learning,
                  {ICML} 2018, Stockholmsm{\"{a}}ssan, Stockholm, Sweden, July
                  10-15, 2018},
  series       = {Proceedings of Machine Learning Research},
  volume       = {80},
  pages        = {3521--3530},
  publisher    = {{PMLR}},
  year         = {2018},
  bibsource    = {dblp computer science bibliography, https://dblp.org}
}

@INPROCEEDINGS {11023255,
author = { Gong, Zirui and Zhang, Yanjun and Zhang, Leo Yu and Zhang, Zhaoxi and Xiang, Yong and Pan, Shirui },
booktitle = { 2025 IEEE Symposium on Security and Privacy (SP) },
title = {{ Not All Edges are Equally Robust: Evaluating the Robustness of Ranking-Based Federated Learning }},
year = {2025},
volume = {},
ISSN = {},
pages = {2527-2545},
doi = {10.1109/SP61157.2025.00251},
publisher = {IEEE Computer Society},
address = {Los Alamitos, CA, USA},
month =May}

@inproceedings{10.5555/3294771.3294783,
author = {Blanchard, Peva and El Mhamdi, El Mahdi and Guerraoui, Rachid and Stainer, Julien},
title = {Machine learning with adversaries: {B}yzantine tolerant gradient descent},
year = {2017},
isbn = {9781510860964},
publisher = {Curran Associates Inc.},
address = {Red Hook, NY, USA},
booktitle = {Proceedings of the 31st International Conference on Neural Information Processing Systems},
pages = {118–128},
numpages = {11},
location = {Long Beach, California, USA},
series = {NIPS'17}
}

@inproceedings{10.1145/3709023.3737691,
author = {Cambus, Melanie and Melnyk, Darya and Milentijevic, Tijana and Schmid, Stefan},
title = {Coordinate-Wise Median in {B}yzantine Federated Learning},
year = {2025},
isbn = {9798400714207},
publisher = {Association for Computing Machinery},
address = {New York, NY, USA},
doi = {10.1145/3709023.3737691},
booktitle = {Proceedings of the International Workshop on Secure and Efficient Federated Learning},
articleno = {5},
numpages = {6},
location = {
},
series = {FL-AsiaCCS '25}
}

@inproceedings{10.1007/978-3-031-25158-0_19,
author = {Xie, Ming and Ma, Jie and Long, Guodong and Zhang, Chengqi},
title = {Robust Clustered Federated Learning with Bootstrap Median-of-Means},
year = {2022},
isbn = {978-3-031-25157-3},
publisher = {Springer-Verlag},
address = {Berlin, Heidelberg},
doi = {10.1007/978-3-031-25158-0_19},
booktitle = {Web and Big Data: 6th International Joint Conference, APWeb-WAIM 2022, Nanjing, China, November 25–27, 2022, Proceedings, Part I},
pages = {237–250},
numpages = {14},
location = {Nanjing, China}
}

@inproceedings{10.1609/aaai.v37i7.26083,
author = {Lyu, Xiaoting and Han, Yufei and Wang, Wei and Liu, Jingkai and Wang, Bin and Liu, Jiqiang and Zhang, Xiangliang},
title = {Poisoning with {C}erberus: stealthy and colluded backdoor attack against federated learning},
year = {2023},
isbn = {978-1-57735-880-0},
publisher = {AAAI Press},
doi = {10.1609/aaai.v37i7.26083},
booktitle = {Proceedings of the Thirty-Seventh AAAI Conference on Artificial Intelligence and Thirty-Fifth Conference on Innovative Applications of Artificial Intelligence and Thirteenth Symposium on Educational Advances in Artificial Intelligence},
articleno = {1014},
numpages = {9},
series = {AAAI'23/IAAI'23/EAAI'23}
}

@INPROCEEDINGS{TranBigData2025,
author= "Duc  A. Tran and Dung Truong and Duy Le",
title= "{RobustFSM}: Submodular Maximization in
Federated Setting with Malicious Clients",
booktitle= "Proceedings of the IEEE International Conference on Big Data (IEEE BigData 2025)",
address="Macao - China",
month="December",
year="2025",
bib2html_dl_pdf = "", 
bib2html_pubtype = {Conference Proceedings},
bib2html_rescat = {Network Theory,Federated Learning, Blockchain}
}

@INPROCEEDINGS{TranBigData2024,
author= "Duong H. Nguyen and Phi L. Nguyen and Truong T. Nguyen and Hieu H. Pham and Duc  A. Tran",
title= "{FedBlock}: A Blockchain Approach to Federated Learning against Backdoor Attacks",
booktitle= "Proceedings of the IEEE International Conference on Big Data (IEEE BigData 2024)",
address="Washington DC",
month="December",
year="2024",
bib2html_dl_pdf = "papers/bigdata2024.pdf", 
bib2html_pubtype = {Conference Proceedings},
bib2html_rescat = {Network Theory,Federated Learning, Blockchain}
}

\end{document}